\documentclass[aps,twocolumn,preprintnumbers,showpacs,showkeys,nofootinbib]{revtex4-1}
\usepackage{epsfig,float,amssymb,latexsym,amsmath,amsthm,fancyhdr}
\usepackage{graphics,longtable}
\newcommand{\ba}{\begin{eqnarray}}
\newcommand{\ea}{\end{eqnarray}}
\newcommand{\be}{\begin{equation}}
\newcommand{\ee}{\end{equation}}
\begin{document}
\author{J.~M.~Alarc\'on}
\email{alarcon@jlab.org}
\affiliation{Theory Center, Jefferson Lab, Newport News, VA 23606, USA}
\author{C.~Weiss} 
\email{weiss@jlab.org}
\affiliation{Theory Center, Jefferson Lab, Newport News, VA 23606, USA}
\preprint{JLAB-THY-17-2525}
\title{Nucleon form factors in dispersively improved Chiral Effective Field Theory I: \\
Scalar form factor}
\begin{abstract}
We propose a method for calculating the nucleon form factors (FFs) of $G$-parity-even operators by
combining Chiral Effective Field Theory ($\chi$EFT) and dispersion analysis. The FFs are 
expressed as dispersive integrals over the two-pion cut at $t > 4 M_\pi^2$. The spectral 
functions are obtained from the elastic unitarity condition and expressed as products of the complex 
$\pi\pi \rightarrow N\bar N$ partial-wave amplitudes and the timelike pion FF. $\chi$EFT is used 
to calculate the ratio of the partial-wave amplitudes and the pion FF, which is real and 
free of $\pi\pi$ rescattering in the $t$-channel ($N/D$ method). The rescattering effects are then 
incorporated by multiplying with the squared modulus of the empirical pion FF. 
The procedure results in a marked improvement compared to conventional $\chi$EFT calculations of the 
spectral functions. We apply the method to the nucleon scalar FF and compute the scalar 
spectral function, the scalar radius, the $t$-dependent FF, and the Cheng-Dashen discrepancy. 
Higher-order chiral corrections are estimated through the $\pi N$ low-energy constants. 
Results are in excellent agreement with dispersion-theoretical calculations. We elaborate
several other interesting aspects of our method. The results show proper scaling behavior in the 
large-$N_c$ limit of QCD because the $\chi$EFT calculation includes $N$ and $\Delta$ intermediate states. 
The squared modulus of the timelike pion FF required by our method can be extracted from 
Lattice QCD calculations of vacuum correlation functions of the operator at large Euclidean distances. 
Our method can be applied to the nucleon FFs of other operators of interest, such as the 
isovector-vector current, the energy-momentum tensor, and twist-2 QCD operators (moments of 
generalized parton distributions).
\end{abstract}
\maketitle
\tableofcontents
\section{Introduction}
\subsection{Form factors and dispersion relations}
Form factors (FFs) are the most basic expressions of the nucleon's complex internal structure and 
finite spatial extent. They parametrize the transition matrix elements of local operators between
nucleon states with different momenta and can be related to the spatial distribution of the 
corresponding physical quantities in localized nucleon states \cite{Miller:2010nz,Burkardt:2002hr}.
The most widely studied FFs are those of the conserved vector and axial vector currents (spin-1 operators),
which describe the interaction of the nucleon with electromagnetic and weak external fields. 
The nucleon vector FFs are measured in elastic 
electron scattering experiments and generally known well \cite{Perdrisat:2006hj}; on the 
axial FFs limited information is available from neutrino scattering and other sources \cite{Bernard:2001rs}.
Besides the conserved currents, there are many more local operators of interest for nucleon structure 
in the context of QCD. The quark and gluon scalar operators (spin-0 operators)
represent the trace of the QCD energy-momentum tensor and measure the contribution of quark 
and gluon fields to the nucleon mass; they also govern the coupling of the nucleon to the 
Higgs boson \cite{Shifman:1978zn}. The corresponding rank-2 traceless tensor operators (spin-2 operators) 
represent the traceless part of the QCD energy-momentum tensor and measure the momentum and angular momentum 
of quarks and gluons in the nucleon, as well as the forces acting on 
them \cite{Jaffe:1989jz,Ji:1996ek,Polyakov:2002yz}.
A much larger class of local QCD operators (spin-$n$ operators, $n \geq 1$) emerges in the 
QCD factorization of hard exclusive processes on the nucleon, in connection with the moments of the 
generalized parton distributions; see Refs.~\cite{Goeke:2001tz,Diehl:2003ny,Belitsky:2005qn,Boffi:2007yc} 
for a review. Because all these operators couple to external fields that are not easily
excited through scattering processes, little is known about the FFs from present experiments. 
It is therefore necessary to develop theoretical methods for calculating the nucleon FFs
of such operators from first principles.

Dispersion relations have proven to be a useful tool in the theoretical analysis of nucleon
FFs. They rely on the analytic properties of the FFs as functions of the invariant momentum
transfer $t$ and connect their behavior in the spacelike and timelike regions, $t < 0$ and $t > 0$.
The FFs are represented as dispersive integrals over their cuts in timelike region, which
describe processes in which the operator couples to the nucleon through exchange of a 
hadronic system in the $t$-channel. For $G$-parity-even operators the hadronic state with the 
lowest mass is the $\pi\pi$ state, and the cut starts is at $t > 4 M_\pi^2$ (two-pion cut).
Examples of such operators are the isovector-vector current, and the isoscalar-scalar and 
isoscalar-spin-2 operators. To evaluate the dispersive integrals one needs to know the
imaginary part of the FFs on the cut (spectral functions). The two-pion cut lies in the
unphysical region below the $N\bar N$ threshold, where the spectral functions cannot be
obtained from timelike nucleon FF data. In the case of the vector and scalar FFs the spectral 
functions on the two-pion cut have been determined using amplitude analysis techniques
with empirical input (unitarity relations with $\pi N$ and $\pi\pi$ scattering 
data \cite{Frazer:1960zza,Frazer:1960zzb,Hohler:1974ht,Belushkin:2005ds,Gasser:1990ap}; 
Roy-Steiner equations \cite{Hoferichter:2016duk,Hoferichter:2012wf}). 
In order to make the dispersive method predictive, and to extend it 
to other operators of interest, one needs a theoretical method to calculate the spectral 
functions of the nucleon FFs.

Chiral Effective Field Theory ($\chi$EFT) represents a systematic method for describing pion 
and nucleon structure and interactions in the low-energy, large-distance regime of strong 
interactions \cite{Gasser:1983yg,Gasser:1984gg}; 
see Refs.~\cite{Bernard:1995dp,Scherer:2002tk,Scherer:2012zzd} for a review. 
It is based on the effective dynamics resulting from the spontaneous breaking 
of chiral symmetry and allows one to calculate amplitudes at pion momenta $k_\pi \sim M_\pi$
in an expansion in $M_\pi / \Lambda_\chi$ with controlled accuracy ($\Lambda_\chi \sim 
1\, \textrm{GeV}$ represents the chiral symmetry breaking scale). The method has been applied 
to the spectral functions of the nucleon FFs on the two-pion cut, using either the 
relativistic or the heavy-baryon formulation for the nucleon 
degrees of freedom \cite{Gasser:1987rb,Bernard:1996cc,Becher:1999he,Kubis:2000zd,Kaiser:2003qp}.
The $\chi$EFT results reproduce the empirical isovector-vector spectral functions at energies
very near the two-pion threshold, $t - 4 M_\pi^2 = \textrm{few} \, M_\pi^2$,
but significantly underestimate the latter at larger energies $t \sim 10$--50 $M_\pi^2$;
see Ref.~\cite{Kaiser:2003qp} for an explicit comparison. The reason for the discrepancy is 
the strong $\pi\pi$ rescattering in the $t$-channel, which manifests
itself in the $\rho$ resonance at $t = 40 \, M_\pi^2 = 0.77\, \textrm{GeV}^2$ and results in an
enhancement of the empirical spectral function. In $\chi$EFT this effect is encoded in 
higher-order $\pi\pi NN$ contact terms and pion loops and would appear in the form of large 
higher-order corrections, which makes the method impractical. A similar situation is observed 
in the spectral function of the scalar FF, where $\pi\pi$ rescattering does not produce a narrow 
resonance but is nevertheless strong. The limited reach of the $\chi$EFT calculations of the 
spectral functions precludes evaluation of the dispersion integral for the FFs based
on $\chi$EFT input alone, as the contributions from larger $t$ require separate modeling. 
In order to extend the reach of $\chi$EFT calculations of the spectral functions beyond 
the near-threshold region one must find a way to account for $\pi\pi$ rescattering in the 
$t$-channel more effectively.

In this article we describe a new method that allows one to construct the spectral functions 
of FFs on the two-pion cut up to larger values of $t$ (in practice, $t \lesssim 1\,\textrm{GeV}^2$)
and enables predictive dispersive calculations of the full nucleon FFs based on $\chi$EFT input alone.
It uses the elastic unitarity condition for the $\pi\pi$ system in the 
$t$-channel \cite{Frazer:1960zza,Frazer:1960zzb} and the $N/D$ method of amplitude 
analysis \cite{Chew:1960iv}. The spectral function of the nucleon FF on the $\pi\pi$ cut 
is expressed as the product of the $\pi\pi \rightarrow N\bar N$ $t$-channel partial-wave amplitude
(PWA) and the complex-conjugate timelike pion FF. The two complex functions have same phase 
on the two-pion cut (Watson theorem) \cite{Watson:1954uc}. $\chi$EFT is used to calculate the 
ratio of the PWA and the timelike pion FF, which is real (it has no two-pion cut) 
and is free of $\pi\pi$ rescattering effects. This function is then multiplied with the squared
modulus of the empirical timelike pion FF, which contains the full $\pi\pi$ rescattering effects. 
The method results in a marked improvement compared to conventional ``direct'' calculations of the 
spectral functions in $\chi$EFT. Realistic spectral functions with controlled uncertainties are 
obtained up to $t \lesssim 1\,\textrm{GeV}^2$. The basic idea was introduced in Ref.~\cite{Alarcon:2017asr} 
in the context of a $\chi$EFT calculation of the nucleon's peripheral transverse densities 
(the Fourier transforms of the FFs) in the LO approximation. Here we describe the general method,
include NLO chiral corrections (fixing of low-energy constants or LECs, convergence, uncertainty estimates), 
and demonstrate the potential for dispersive calculations of the FFs proper and their 
derivatives, which opens up a range of new applications. We also explore other interesting
aspects of the new method. We point out a possible combination with first-principles 
calculations of the squared modulus of the timelike pion FF with Euclidean correlation functions 
(e.g.\ Lattice QCD), which could be used for the dispersive calculation of FFs of QCD 
operators whose pion FF are not known empirically. We demonstrate that our nucleon FF 
results have the correct scaling behavior in the large-$N_c$ limit of QCD because the 
$\chi$EFT amplitudes include $N$ and $\Delta$ intermediate states.

Here we apply the method to the nucleon scalar FF and its spectral function. The choice is motivated 
by pedagogical considerations and physical interest. The scalar density is the simplest operator,
with only a single nucleon FF, and a single $t$-channel partial wave in the unitarity relation
for the spectral function.
The pion scalar FF has been determined from dispersion analysis with $\chi$EFT constraints and 
is available as input for our calculation \cite{Ananthanarayan:2004xy,Oller:2007xd,Celis:2013xja}. 
The scalar nucleon FF has been computed using amplitude analysis techniques 
and serves as a reference point for our results \cite{Gasser:1990ap,Hoferichter:2012wf}. 
The scalar FF thus represents the ideal testing ground for our method. It is also of great physical 
interest in itself, in connection with the nucleon mass problem and the coupling to the scalar
sector of the standard model (see below). Applications of our method to the nucleon
isovector-vector FFs will be presented in a forthcoming article \cite{Alarcon:vector}.

The plan of this article is as follows. In the remainder of this section we summarize the basic 
properties of the scalar FF and its dispersive representation. In Sec.~\ref{sec:method} we describe
the general method of dispersively improved $\chi$EFT, including the elastic unitarity relation and 
$N/D$ method, LO $\chi$EFT calculations, estimates of higher-order corrections, and the properties of the
pion FF. In Sec.~\ref{sec:results} we apply the method to the nucleon scalar 
spectral function and use it to calculate the nucleon scalar radius, the scalar FF, and
the Cheng-Dashen discrepancy. In Sec.~\ref{sec:discussion} we discuss the extraction of the
timelike pion FF from Euclidean correlation functions and the correspondence of our approach
with large-$N_c$ QCD. An outlook on further applications is presented in Sec.~\ref{sec:outlook}.

A combination of $\chi$EFT and dispersion theory similar to the one used here was proposed 
in the context of a recent study of hyperon transition FFs \cite{Granados:2017cib}. 
Techniques related to the $N/D$ method were also applied in earlier $\chi$EFT studies of 
meson-meson, meson-baryon, and baryon-baryon 
scattering \cite{Alarcon:2011kh,Oller:1998zr,Oller:2000ma,Meissner:1999vr,Oller:2000fj,%
Albaladejo:2011bu,Albaladejo:2012sa}.  
\subsection{Scalar form factor}
The scalar density of light quarks in QCD is measured by the composite local operator
\be
O_\sigma(x) \; \equiv \; \hat m \; \sum_{f = u, d} \bar\psi_f(x) \psi_f(x) ,
\label{operator_def}
\ee
where $\psi_f(x) \; (f = u, d)$ is the quark field and $\hat m \equiv m_u = m_d$ the quark
mass (we assume isospin symmetry). The operator Eq.~(\ref{operator_def}) is scale-independent 
and represents the quark mass term in the QCD Lagrangian and Hamiltonian densities. The same
operator appears in the trace of the QCD energy-momentum tensor, alongside the gluonic and 
strange-quark scalar operators and a similar light-quark operator that results 
from the trace anomaly; see Ref.~\cite{Shifman:1978zn} for details.
The transition matrix element of the operator Eq.~(\ref{operator_def}) between nucleon states with
4-momenta $p$ and $p'$ is of the form
\be
\langle N(p', s') | \, O_\sigma (0) \, | N(p, s)\rangle =  \bar u' u \; \sigma(t) ,
\label{scalar_def}
\ee
where $\bar u'$ and $u$ are the nucleon 4-spinors, and $\sigma(t)$ is the nucleon scalar FF. 
It is a function of the invariant momentum transfer $t \equiv (p' - p)^2$,
with $t < 0$ in the physical region of the nucleon transition (spacelike FF). The corresponding
timelike FF is defined analogously, as the matrix element between the vacuum and a 
nucleon-antinucleon state, with $t \equiv (p' + p)^2 > 0$. The matrix elements are
diagonal in isospin ($N =$ proton, neutron).

The scalar FF is an analytic function of $t$. The physical sheet has cuts along the
positive real axis, which result from processes in which the operator creates a
hadronic state that couples to the $N\bar N$ system,
\be
\textrm{operator} \rightarrow \textrm{hadronic state} \rightarrow N\bar N ;
\label{tchannel_process}
\ee
such processes occur in the unphysical region below the $N\bar N$ threshold. 
The lowest-mass hadronic state with scalar quantum numbers is the $\pi\pi$ state
with threshold at $t = 4 M_\pi^2$ (two-pion cut); other hadronic states ($4\pi$ etc.)
give rise to further cuts with higher thresholds; the cuts can be combined to a
principal cut starting at $t = 4 M_\pi^2$. One can thus write dispersion relations
that express the FF in the complex plane as an integral over the discontinuity on
the principal cut. In practice one considers a once-subtracted dispersion relation,
\be
\sigma(t) \; = \; \sigma(0) + \frac{t}{\pi} \int_{4M_\pi^2}^\infty dt' \; 
\frac{\text{Im}\,\sigma(t')}{t'(t'-t)} ,
\label{ff_dispersive}
\ee
which suppresses contributions from large $t'$ and ensures rapid convergence of 
the integral (see below). It determines the FF up to a subtraction constant, 
which is chosen as the value of the FF at $t = 0$, $\sigma(0)$, the so-called
pion-nucleon sigma term. The integration is over the imaginary part of the FF on
the principal cut, $\textrm{Im}\, \sigma(t')$, which is referred to as the spectral
function.

Of particular interest is the behavior of the scalar FF near $t = 0$. The
derivative of the FF at $t = 0$ defines nucleon's scalar charge radius,
\be
\langle r^2 \rangle_\sigma \; \equiv \; \left.\frac{6}{\sigma(0)}\frac{d\sigma}{dt}\right|_{t=0} .
\ee
The finite difference
\be
\Delta_\sigma \; \equiv \; \sigma(t = 2M_\pi^2) - \sigma(t = 0)
\label{Delta}
\ee
is needed in the extraction of the sigma term from $\pi N$ scattering data using the
Cheng-Dashen theorem \cite{Cheng:1970mx}, which connects the Born-subtracted isoscalar 
$\pi N$ scattering amplitude at $s=m_N^2$ and $t=2M_\pi^2$ to $\sigma (t = 2M_\pi^2)$. 
The dispersive representation of these quantities is
\ba
\langle r^2 \rangle_\sigma &=& \frac{6}{\pi \sigma(0)} \int_{4M_\pi^2}^\infty dt' \;
\frac{\text{Im}\,\sigma(t')}{t'^2} ,
\label{r2_dispersive}
\\[1ex]
\Delta_\sigma &=& \frac{2 M_\pi^2}{\pi} \int_{4M_\pi^2}^\infty dt' \;
\frac{\text{Im}\,\sigma(t')}{t'(t' - 2 M_\pi^2)} .
\label{Delta_dispersive}
\ea
The convergence of these integrals at large $t'$ is similar to that of the once-subtracted 
dispersion relation for the FF, Eq.~(\ref{ff_dispersive}).

\begin{figure}
\begin{center}
\epsfig{file=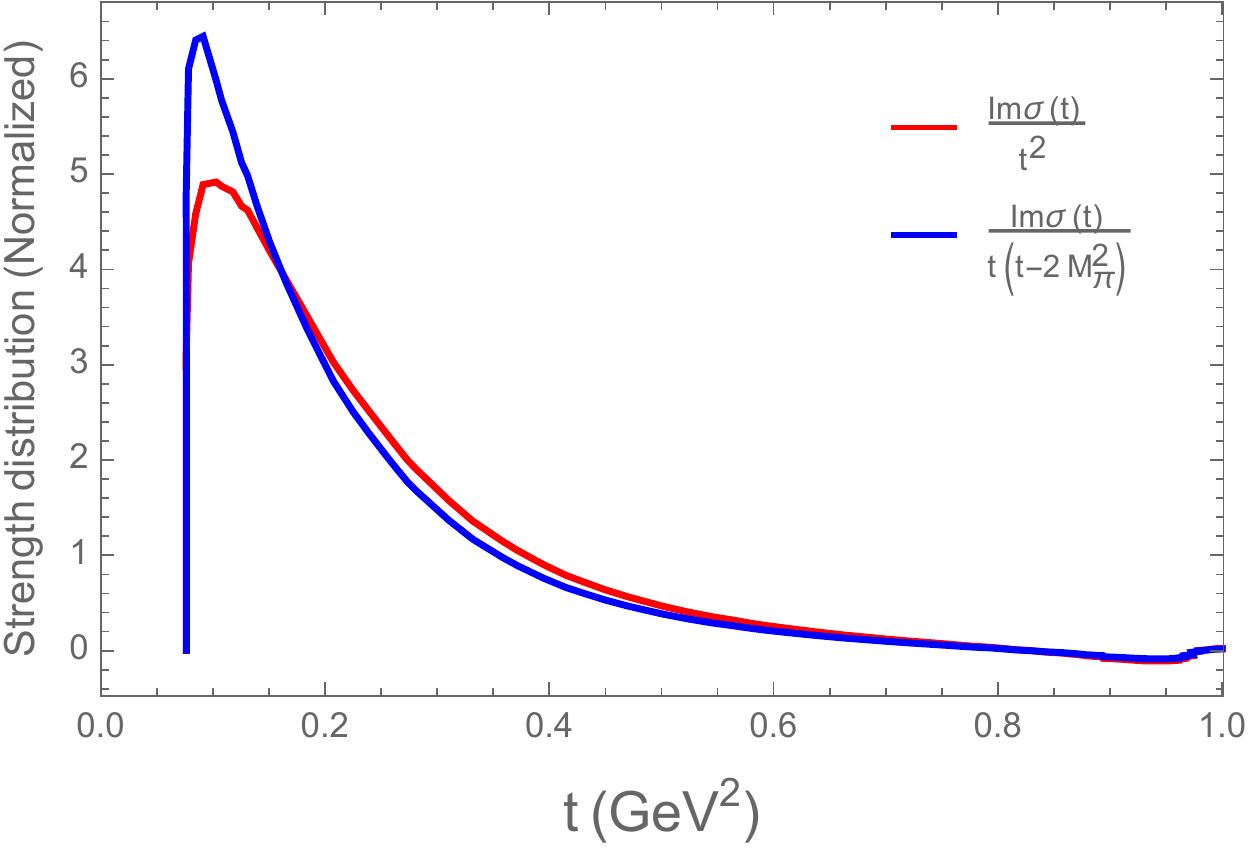,width=.45\textwidth,angle=0}
\caption{\small Distribution of strength in the dispersive integrals for the scalar charge radius,
Eq.~(\ref{r2_dispersive}), and the Cheng-Dashen discrepancy, Eq.~(\ref{Delta_dispersive}).
The plot shows the integrands as functions of $t$, divided by the value of the integral, i.e., 
normalized to unit area under the curves.
\label{fig:strength}}
\end{center}
\end{figure}

The spectral function of the scalar nucleon FF has been constructed using amplitude analysis 
techniques with empirical input \cite{Gasser:1990ap,Hoferichter:2012wf}. 
Figure~\ref{fig:strength} shows the distribution of strength in the dispersive integrals 
Eqs.~(\ref{r2_dispersive}) and (\ref{Delta_dispersive}). One sees that the integral
converges rapidly, and that the main contribution comes from the region $t' \lesssim 
0.5\, \textrm{GeV}^2$. This determines the range where one needs to
calculate spectral function if one aims for a first-principles calculation of the
scalar quantities through their dispersive integrals.

Evaluation of the integrals with the empirical spectral functions of Ref.~\cite{Gasser:1990ap} 
has found $\langle r^2 \rangle \sim 1.6 \, \textrm{fm}^2$, substantially larger
than the proton's charge radius $\langle r^2 \rangle_1 \sim 0.65 \, \textrm{fm}^2$ (Dirac radius).
The discrepancy $\Delta_\sigma$ has been obtained at $\sim 14\, \textrm{MeV}$. 
The significance of these findings will be discussed in Sec.~\ref{sec:results}.

The scalar FF of the pion is defined analogously to that of the nucleon
in Eq.~(\ref{scalar_def}),
\be
\langle \pi(p') | \, O_\sigma(0) \, | \pi(p)\rangle \; = \; \sigma_\pi (t) ,
\label{scalar_pion}
\ee
where $\pi = \pi^+, \pi^-, \pi^0$ (isospin symmetry) and $t = (p' - p)^2 < 0$
in the physical region. The value at $t = 0$ is
\be
\sigma_\pi (0) \;\; = \;\; M_\pi^2 ,
\ee
which follows from the fact that the scalar operator corresponds to the 
chiral-symmetry-breaking pion mass term in the chiral Lagrangian.
The corresponding timelike FF is defined as
\be
\langle 0 | \, O_\sigma(0) \, | \pi(p') \pi(p) \rangle \; = \; \sigma_\pi (t) ,
\label{ff_pion_timelike}
\ee
where now $t = (p + p')^2 > 4 M_\pi^2$ in the physical region. The scalar FF
of the pion is of physical interest in itself, and enters in dispersive calculations 
of nucleon scalar FF.
\section{Method}
\label{sec:method}
\subsection{Dispersively improved $\chi$EFT}
\begin{figure}
\begin{center}
\includegraphics[width=.48\textwidth]{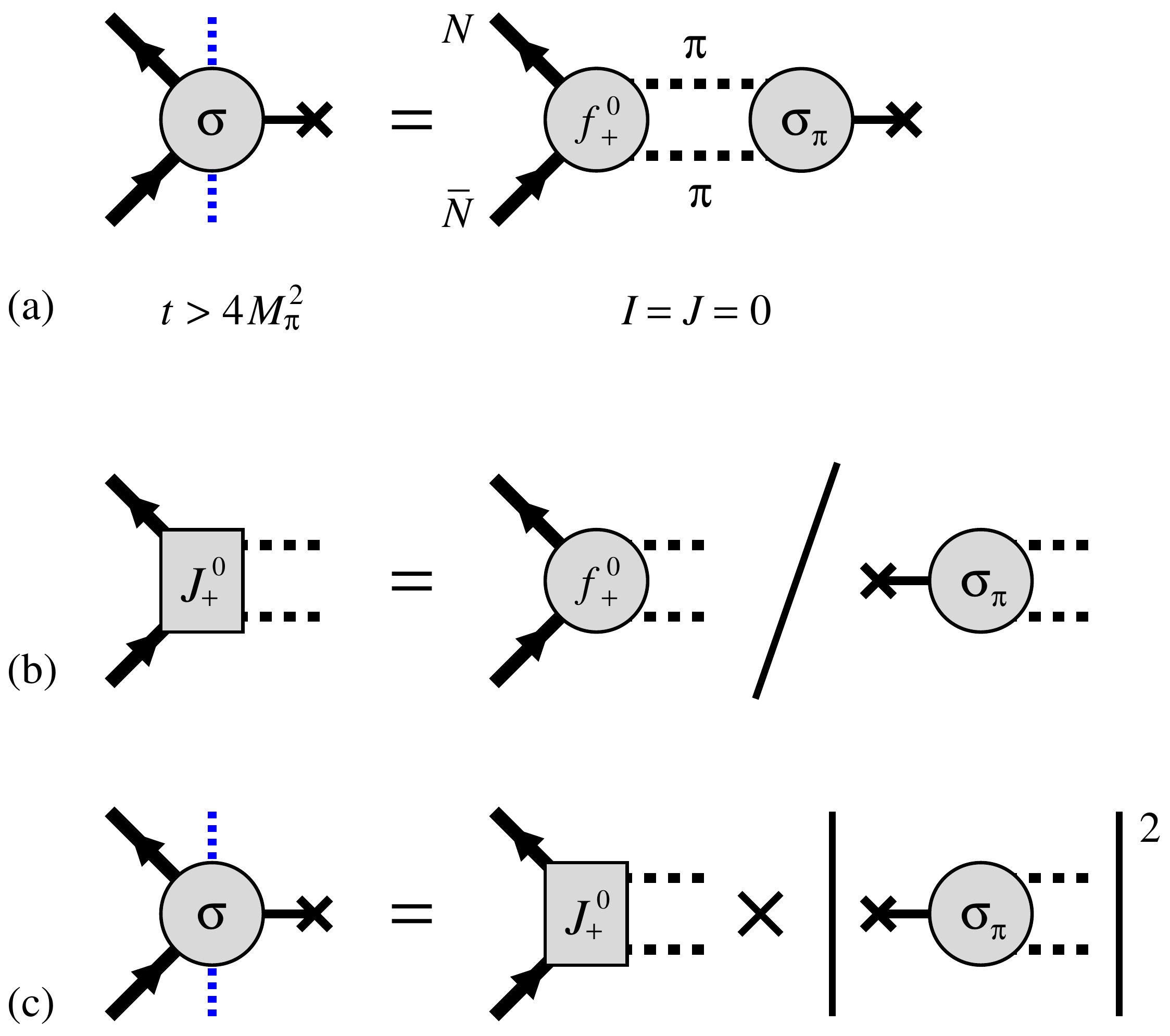}
\caption{\small (a)~Unitarity relation for the imaginary part of the nucleon scalar FF
on the two-pion cut, Eq.~(\ref{unitarity}). (b) Real function $J_+^0(t)$, Eq.~(\ref{J_def}),
defined as the ratio of the $\pi\pi \rightarrow N\bar N$ PWA and the pion FF. 
(c)~Unitarity relation in terms of $J_+^0(t)$ and the squared modulus
of the pion FF, Eq.(\ref{unitarity_real}).
\label{fig:unitarity}}
\end{center}
\end{figure} 
We now describe the method for calculating the spectral function of nucleon FFs on the two-pion cut 
in $\chi$EFT using a representation based on the elastic unitarity condition and the $N/D$ method.
While we use the scalar FF as a specific example, the method is general and can be applied to 
the FFs of any $G$-parity-even operator coupling to the $\pi\pi$ state.

\begin{figure}
\begin{center}
\includegraphics[width=.36\textwidth]{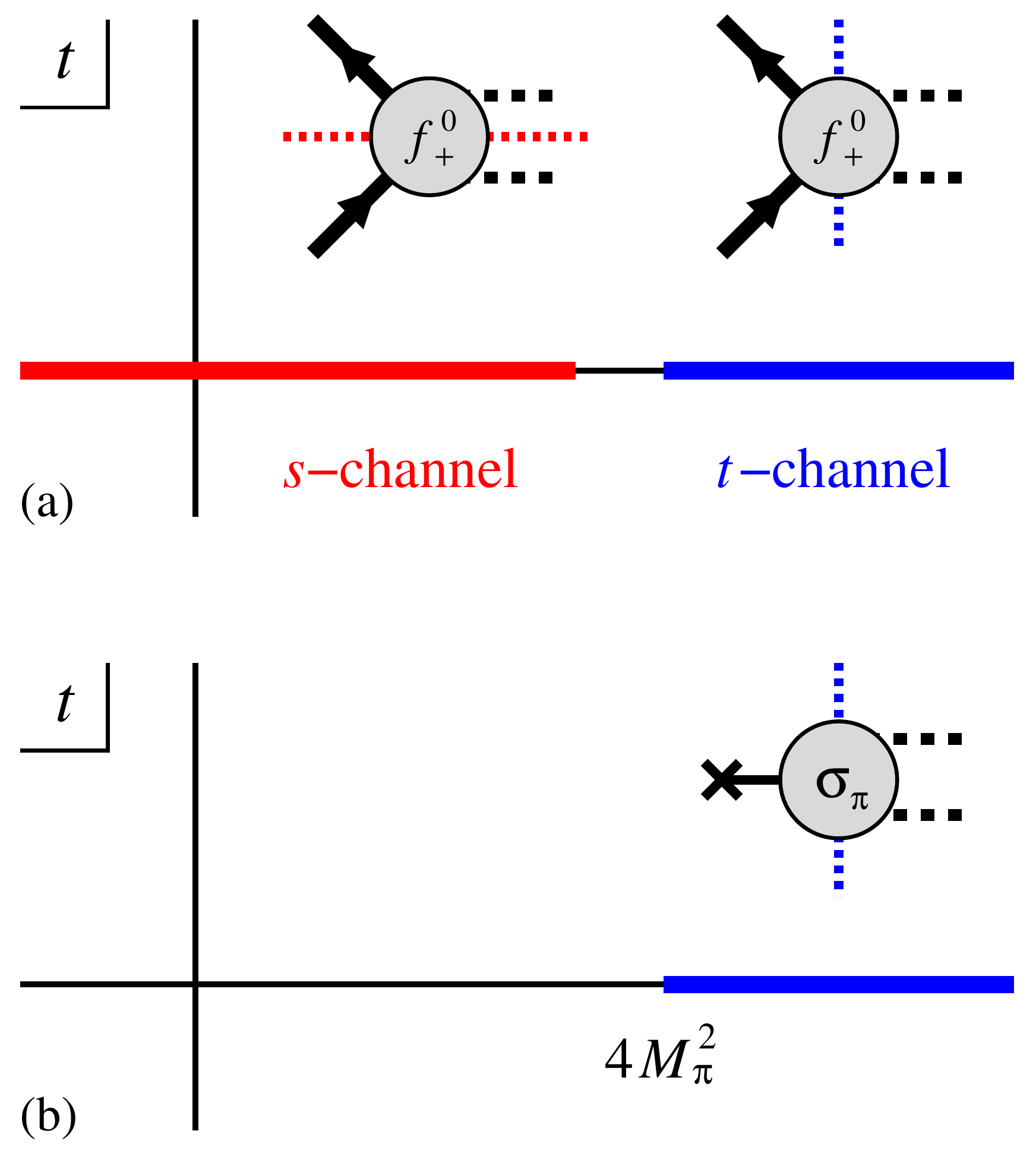}
\caption{\small (a)~Analytic structure of the $\pi\pi \rightarrow N\bar N$ PWA.
The function has a right-hand cut resulting from the $t$-channel $\pi\pi$ intermediate state, 
and a left-hand cut resulting from $s$-channel intermediate states ($N, \Delta, \pi N$, \ldots).  
The phase of the PWA on the right-hand cut is the same as that of the pion FF. 
(b)~Analytic structure of the pion FF. The function has a right-hand cut 
resulting from the $\pi\pi$ intermediate state.
\label{fig:cuts}}
\end{center}
\end{figure}

In the region $4 M_\pi^2 < t < 16 M_\pi^2$ only the $\pi\pi$ state contributes to the discontinuity 
of the nucleon FF through the process Eq.~(\ref{tchannel_process}). In this situation the spectral function 
can be computed using the elastic unitarity condition, which expresses the conservation of flux 
in the $t$-channel \cite{Frazer:1960zza,Frazer:1960zzb,Hohler:1974ht}. For the scalar nucleon FF it 
takes the form \cite{Gasser:1990ap}
\be
\text{Im} \, \sigma(t) \; = \; \frac{3 k_{\rm cm}}{4 \widetilde p_N^2 \sqrt{t}} \; f^0_+(t) \,
\sigma_\pi^*(t) ,
\label{unitarity}
\ee
where 
\be
k_{\rm cm} \equiv \sqrt{t/4 - M_\pi^2}
\label{k_cm}
\ee
is the center-of-mass momentum of the pions in the $\pi\pi$ system, 
and
\be
\widetilde p_N \equiv {\textstyle \sqrt{m_N^2 - t/4}}
\label{p_cm}
\ee
is related to the unphysical momentum of the nucleons in the $N\bar N$ system (see Fig.~\ref{fig:unitarity}a). 
The function $f^0_+(t)$ is the $I=J=0$ $\pi\pi \rightarrow N \bar N$ $t$-channel PWA, and $\sigma^\ast_{\pi}(t)$ 
is the complex conjugate of the timelike pion FF Eq.~(\ref{ff_pion_timelike}). 
While the unitarity condition applies at real $t > 4 M_\pi^2$ on the upper edge of the cut
($t \rightarrow t + i0$), the functions $f^0_+(t)$ and $\sigma_{\pi}(t)$ are 
defined for arbitrary complex $t$, and it is worthwhile to recall their analytic structure.
The PWA $f^0_+(t)$ has both a right-hand cut and a left-hand cut 
(see Fig.~\ref{fig:cuts}a). The right-hand cut results from $t$-channel processes
with the $\pi\pi$ intermediate state and starts at $t = 4 M_\pi^2$. The left-hand cut results 
from $s$-channel processes with intermediate baryonic states ($N, \Delta, \pi N$, \ldots ) and starts 
at $t = 4 M_\pi^2 - M_\pi^4/m_N^2$ for the intermediate $N$. The two cuts are thus of
different physical origin. The pion FF $\sigma(t)$ has only a right-hand cut starting at 
$t = 4 M_\pi^2$, resulting from the $\pi\pi$ intermediate state, which is just the two-pion 
cut of the pion FF (see Fig.~\ref{fig:cuts}b).

A crucial point is that the complex functions $f^0_+(t)$ and $\sigma_{\pi}(t)$ have the same
phase on the right-hand cut (the two-pion cut). Physically, this follows from the
fact that the phases of the two amplitudes arise from the same elastic $\pi\pi$ rescattering 
processes (Watson theorem) \cite{Watson:1954uc}. 
Mathematically, this is necessary for the product of $f^0_+(t)$ and $\sigma^\ast_{\pi}(t)$ 
to result in the real function $\textrm{Im}\, \sigma(t)$, as was 
already implied in the unitarity condition Eq.(\ref{unitarity}).
This circumstance allows one to rewrite the unitarity relation in a manifestly
real form \cite{Frazer:1960zza} (see Fig.~\ref{fig:unitarity} b and c)
\ba
\text{Im} \, \sigma(t) &=& \frac{3 k_{\rm cm}}{4 \widetilde p_N^2 \sqrt{t}} \; 
J^0_+(t) \, |\sigma_\pi(t)|^2 ,
\label{unitarity_real}
\\[1ex]
J^0_+(t) &\equiv& \frac{f^0_+(t)}{\sigma_\pi(t)} .
\label{J_def}
\ea
The function $J^0_+(t)$ is real at $t > 4 \, M_\pi^2$ and therefore has no right-hand cut;
if it had one, there would be a discontinuity resulting in a non-zero imaginary part.
It does have a left-hand cut, inherited from the PWA $f^0_+(t)$.
The squared modulus $|\sigma_\pi(t)|^2$ is obviously real. The representation of 
Eqs.~(\ref{unitarity_real}) and (\ref{J_def}) permits a simple physical interpretation.
Since the phase of $f^0_+(t)$ and $\sigma_{\pi}(t)$ arises from $\pi\pi$ rescattering 
processes, the equations effectively separate the $\pi\pi \rightarrow N\bar N$ 
coupling [contained in $J^0_+(t)$, in which the phase cancels] from the $\pi\pi$ 
rescattering [contained in $|\sigma_\pi(t)|^2$, which is a purely pionic amplitude].
This interpretation can provide useful guidance for the following.

The representation of Eqs.~(\ref{unitarity_real}) and (\ref{J_def}) is equivalent to 
applying the $N/D$ method to the 
$\pi\pi \rightarrow N\bar N$ PWA \cite{Chew:1960iv}. In this approach the PWA is 
represented in the form $f^0_+(t) = N(t)/D(t)$, such that the right-hand cut (related 
to the $t$-channel exchanges) appears only in the factor $1/D(t)$, and the left-hand cut 
(related to the $s$-channel intermediate states) appears in the factor $N(t)$. In the case at 
hand the $D$ function is chosen as the inverse pion FF, $D(t) = 1/\sigma_\pi(t)$,
and the $N$ function is given by Eq.~(\ref{J_def}), $N(t) = J^0_+(t)$ \cite{Hohler:1974ht}.

The representation of Eqs.~(\ref{unitarity_real}) and (\ref{J_def}) suggests a new approach 
to calculating the spectral function of nucleon FFs on the two-pion cut in $\chi$EFT. 
We use $\chi$EFT to compute the real function $J^0_+(t)$ at $t > 4 \, M_\pi^2$ 
to a fixed order. We then multiply the result with the empirical $|\sigma_\pi(t)|^2$, which contains 
the effects of $\pi\pi$ rescattering. This approach has several advantages compared to ``direct'' 
calculations of the nucleon spectral functions:
\begin{itemize}
\item[(a)] The $\chi$EFT calculations of $J^0_+(t)$ are not affected by $\pi\pi$ rescattering, 
as the latter is contained entirely in $|\sigma_\pi(t)|^2$. The rescattering effects are strong 
and would require large higher-order corrections when treated within $\chi$EFT. We therefore 
expect the new approach to show much better convergence than direct $\chi$EFT calculations of 
the spectral function. Higher-order corrections can perturbatively improve the coupling of the 
$\pi\pi$ system to the nucleon described by $J^0_+(t)$, while the rescattering effects described
by $|\sigma_\pi(t)|^2$ are taken from other sources (dispersion theory, data, Lattice QCD).
\item[(b)] The organization according to Eqs.~(\ref{unitarity_real}) and (\ref{J_def})
is consistent with the idea of ``separation of scales'' basic to $\chi$EFT. The function $J^0_+(t)$ 
is dominated by the singularities of the $N$ and $\Delta$ Born diagrams, or diagrams with 
$\pi N$ inelastic intermediate states in higher orders, which are governed by the 
scales $M_\pi$ and $m_\Delta - m_N$. The $t$-dependence of the pion FF, in contrast, 
is governed by the chiral-symmetry-breaking scale $\Lambda_\chi \sim 1\, \textrm{GeV}$.
The intrinsic logic of $\chi$EFT therefore suggests to apply the $\chi$EFT calculations to
$J^0_+(t)$ and treat $|\sigma_\pi(t)|^2$ as an external input.
\item[(c)] The timelike pion FF enters only through its squared modulus $|\sigma_\pi(t)|^2$,
not its phase. This reduces model dependence in the determination of the empirical pion FF
and represents an advantage over approaches working with the original unitarity 
condition, Eq.~(\ref{unitarity}), where the pion FF enters as a complex function.
The squared modulus of the timelike scalar pion FF can be extracted directly from Euclidean 
vacuum-to-vacuum correlation functions of the scalar operator, which can be computed in Lattice QCD 
(see Sec.~\ref{subsec:euclidean}). In the electromagnetic case the squared modulus of the 
timelike pion FF can directly be measured in $e^+e^- \rightarrow \pi^+\pi^-$ exclusive annihilation 
experiments.
\end{itemize}
We refer to the new method as ``dispersively improved $\chi$EFT'' (DI$\chi$EFT).
The method is applicable strictly at $4 M_\pi^2 < t < 16\, M_\pi^2$, where only the $\pi\pi$ 
channel is open and the elastic unitarity condition Eq.~(\ref{unitarity}) is valid.
It is expected that inelasticities from other channels ($4\pi$) are small up to the
$K\bar K$ threshold; neglecting those the representation of
Eqs.~(\ref{unitarity_real}) and (\ref{J_def}) can effectively be used up to 
$t \sim 1\, \textrm{GeV}^2$. Our method could thus in principle be applied up
to such values of $t$, provided that the $\chi$EFT calculations of $J^0_+(t)$
converge sufficiently well (this question will be investigated below).
\subsection{Leading-order calculation}
For the calculation of $J^0_+(t)$ we use $SU(2)$-flavor $\chi$EFT with relativistic 
$N$ and $\Delta$ degrees of freedom. The relativistic formulation ensures the correct 
analytic structure of the amplitudes (position of branch points, threshold behavior),
which is critical in the present application. The inclusion of the $\Delta$ as an
explicit degree of freedom is needed because the $\Delta$ Born term makes important
contributions to the $\pi\pi \rightarrow N\bar N$ PWA (see below); it is also needed 
to reproduce the correct scaling behavior of the spectral function in the large-$N_c$ 
limit of QCD (see Sec.~\ref{subsec:largenc}). These features have proved to be essential also 
in other applications of baryon $\chi$EFT to $\pi N$ scattering, photoproduction, 
and nucleon structure \cite{Alarcon:2011zs,Alarcon:2012nr,Alarcon:2013cba,Lensky:2014dda,%
Lensky:2009uv,Blin:2014rpa,Blin:2015era,Blin:2016itn,Ledwig:2011cx,Geng:2008mf,Ledwig:2014rfa}.

The basic setup of the relativistic $\chi$EFT used in the present study 
(fields, Lagrangian, power counting, couplings) is described 
in Ref.~\cite{Alarcon:2012kn} and summarized in Ref.~\cite{Alarcon:2017asr}. 
The spin-1/2 $N$ is described by a relativistic bispinor field (Dirac field). 
The spin-3/2 $\Delta$ is introduced as a 4-vector-bispinor field, 
which has to be subjected to relativistically covariant constraints to eliminate spurious spin-1/2 
degrees of freedom. Here we use the formulation in which the spin-1/2 degrees of 
freedom are allowed to propagate but are filtered out at the interaction vertices 
(consistent vertices) \cite{Pascalutsa:1998pw,Pascalutsa:1999zz, Pascalutsa:2000kd,Krebs:2008zb}.
The construction of the chiral Lagrangian with the spin-3/2 fields has been described 
in Refs.~\cite{Hemmert:1996xg,Hemmert:1997ye}. Several expansion schemes have been proposed
for the $\chi$EFT with the $\Delta$, assuming certain parametric relations between the chiral 
parameters $k_\pi \sim M_\pi$ and the $N$-$\Delta$ mass splitting $m_\Delta - m_N$.
In the present application the differences between the various 
expansion schemes for the $\Delta$ are irrelevant, because the calculations are carried out
at an accuracy where $\Delta$ loops do not enter. The only difference to $\chi$EFT with 
$N$ only is in the appearance of the $\Delta$ Born graphs at leading order. We therefore denote 
the order of our calculations by LO, NLO, N2LO, as is common in $\chi$EFT with $N$ only. 

Regarding the power counting, we note that $\chi$EFT calculations with relativistic baryons must 
in principle deal with power-counting-breaking 
terms arising from chiral loops with baryons, i.e., lower-order terms in chiral counting resulting from 
higher-order terms in the loop expansion. The standard power counting for loops can be recovered 
by adopting the extended-on-mass-shell (EOMS) scheme \cite{Fuchs:2003qc}. While diagrams with chiral 
loops are not considered in the present study, it is important to mention this scheme here, as it ensures 
that the tree-level results are not mixed up with power-counting-breaking terms arising from chiral loops.

%
%
\begin{figure}
\begin{center}
\includegraphics[width=.36\textwidth]{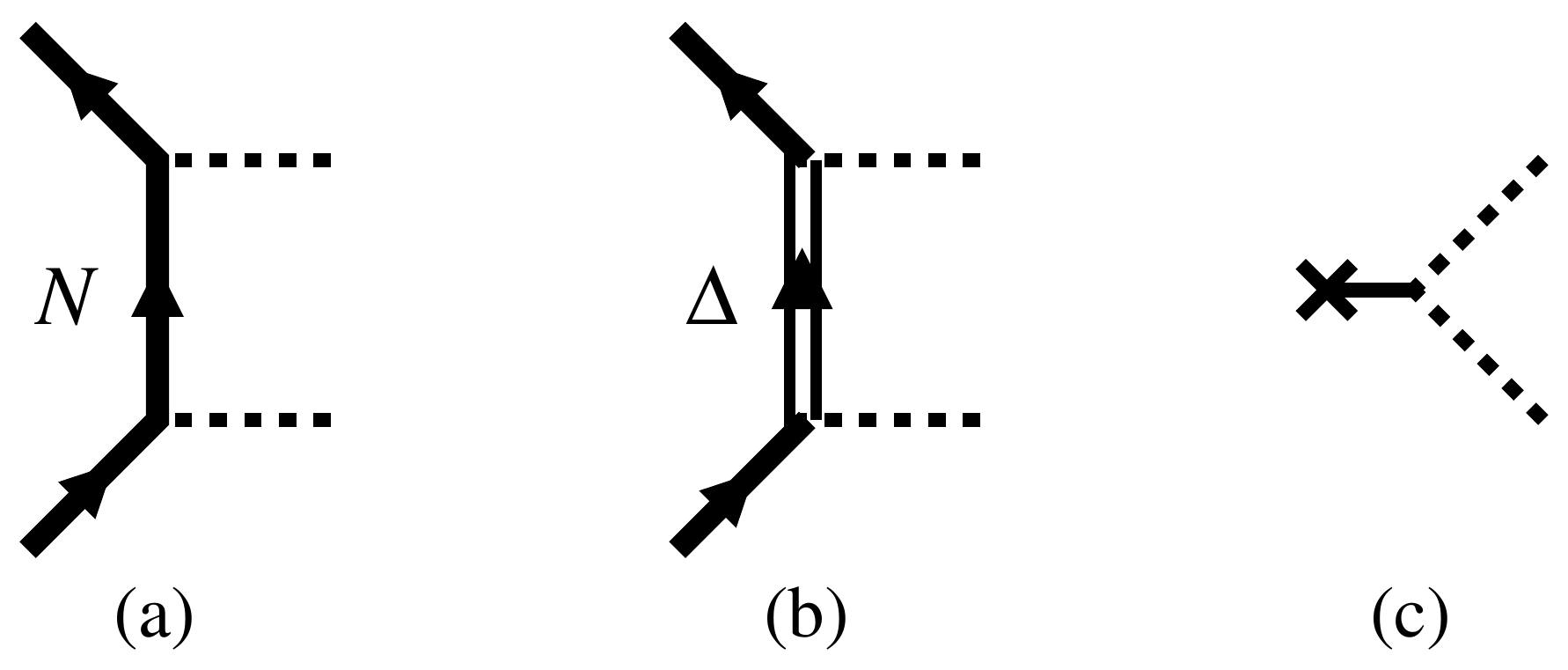}
\caption{\small (a, b) LO $\chi$EFT diagrams contributing to the $\pi\pi \rightarrow N\bar N$
PWA. (c) Pion scalar FF in LO.
\label{fig:eft}}
\end{center}
\end{figure} 
The LO $\chi$EFT diagrams for the $I = J = 0$ $\pi\pi \rightarrow N\bar N$ partial-wave
amplitude $f^0_+(t)$ are the $N$ Born term shown in Fig.~\ref{fig:eft}a and the
$\Delta$ Born term in Fig.~\ref{fig:eft}b; $\pi\pi NN$ contact terms appear
only in higher orders and will be discussed below.
In the LO calculation of the ratio $J^0_+(t)$, Eq.~(\ref{J_def}), the pion FF in the denominator 
is evaluated at LO, see Fig.~\ref{fig:eft}c. At this order in $\chi$EFT the pion is pointlike,
$\sigma_\pi(t) \equiv \sigma_\pi(0) = M_\pi^2$. The LO result for $J^0_+(t)$ is therefore just
the result for $f^0_+(t)$ divided by $M_\pi^2$. At this accuracy our approach based on
Eq.~(\ref{unitarity_real}) simply amounts to multiplying the LO $\chi$EFT result
for the nucleon spectral function $\textrm{Im}\, \sigma(t)$ (as obtained by direct $\chi$EFT 
calculation of the spectral function without the unitarity condition) 
by the normalized empirical pion FF $|\sigma_\pi(t)|^2/M_\pi^4$,
\newline
\be
\textrm{Im} \, \sigma(t) \;\; = \;\;
\textrm{Im} \, \sigma(t) \, [\textrm{LO}] \, \times \, \frac{|\sigma_\pi(t)|^2}{M_\pi^4} .
\label{improvement}
\ee
This formula permits an extremely simple implementation of unitarity at LO accuracy. 
The factor $|\sigma_\pi(t)|^2/M_\pi^4$ describes the enhancement of the direct 
$\chi$EFT result for the spectral function due to $\pi\pi$ rescattering. 
Numerical results obtained with this approximation will be presented below.

%
%
\begin{figure}
\begin{center}
\includegraphics[width=.45\textwidth]{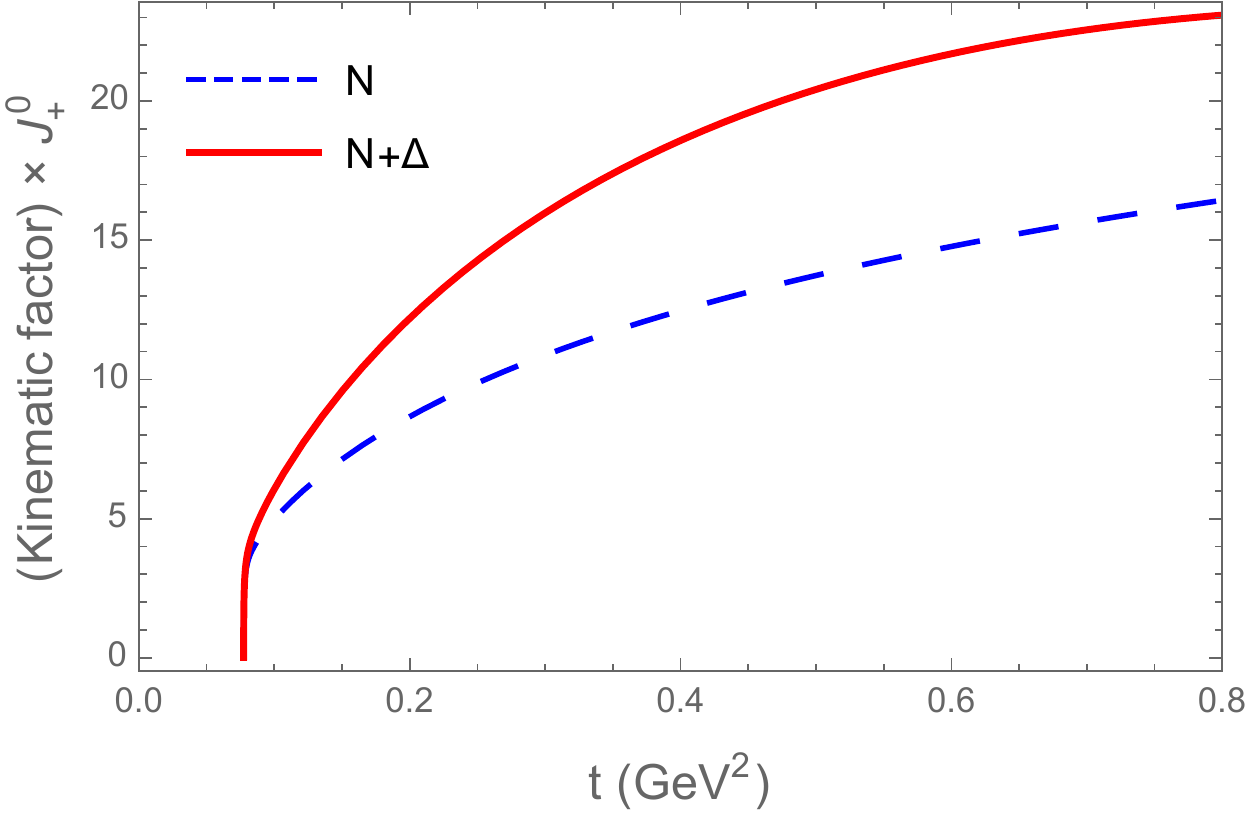}
\caption{\small LO $\chi$EFT result for 
$[3 k_{\rm cm}/(4 \widetilde p_N^2 \sqrt{t})] \, J^0_+(t)$, Eq.~(\ref{unitarity_real}). 
Dashed line: Contribution of $N$ Born term, Fig.~\ref{fig:eft}a. 
Solid line: Sum of $N$ and $\Delta$ Born terms, Fig.~\ref{fig:eft}a and b.
\label{fig:ndelta}}
\end{center}
\end{figure} 
The analytic expressions for the LO $\chi$EFT results for $J^0_+(t)$ are given in 
Appendix~\ref{app:expressions}. The numerical values are shown in Fig.~\ref{fig:ndelta}.
For a better view the plot shows the function multiplied by the kinematic factor of 
Eq.(\ref{unitarity_real}), $3 k_{\rm cm}/(4 \widetilde p_N^2 \sqrt{t}) \, J^0_+(t)$; 
this combination is equal to $\text{Im} \, \sigma(t)/|\sigma_\pi(t)|^2$ by 
virtue of Eq.(\ref{unitarity_real}). One observes that the contributions from the
$N$ and $\Delta$ Born term amplitudes have the same sign and are roughly comparable
in magnitude.
\subsection{Estimates of higher-order corrections}
\label{subsec:higher-order}
At NLO accuracy corrections to the $\pi N$ scattering amplitude arise only from NLO 
$\pi\pi NN$ contact terms in the chiral Lagrangian. The NLO contributions to the
$I = J = 0$ $\pi\pi \rightarrow N\bar N$ PWA in Eq.~(\ref{J_def}) therefore have a 
simple structure. Corrections to the pion FF appear only at N2LO accuracy through pion loops. 
The expression for the NLO corrections to $J_+^0(t)$ is given in Eq.~(\ref{J_NLO}) of
Appendix~\ref{app:expressions}. At this accuracy Eq.~(\ref{improvement}) is still valid,
and the NLO corrections to the spectral function are obtained simply by replacing
$J_+^0(t)$ by its NLO expression.

\begin{figure}[t]
\begin{center}
\includegraphics[width=.42\textwidth]{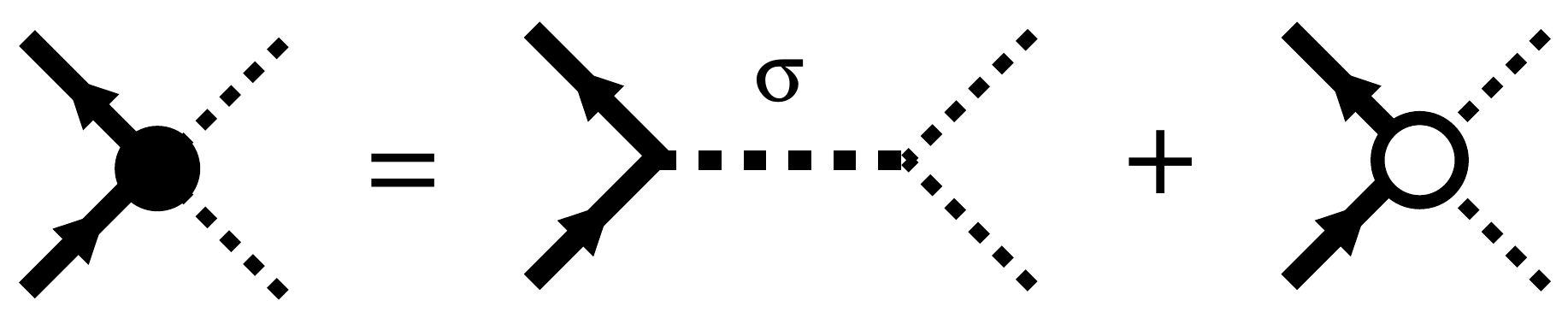}
\caption{\small Adjustment of the LECs of the NLO $\pi\pi NN$ contact term in our unitarity-based
approach. The original contact term (filled circle, left-hand side) is equated
with the sum of $\sigma$ meson exchange and a reduced contact term (open circle).
The reduced contact terms are used in the present scalar FF calculation 
with explicit unitarization.
\label{fig:contact}}
\end{center}
\end{figure}

For evaluating the higher-order corrections we use the LECs of $\pi N$ scattering.
The values have to be adjusted consistently with the logic of our unitarity-based approach.
The LECs in standard $\chi$EFT absorb rescattering effects that are treated explicitly
within our unitarity-based approach. The contact terms appropriate for our approach are 
therefore obtained by subtracting the effects of rescattering from the original LECs.
To do this in practice, we describe the rescattering effects in the $I = J = 0$ $\pi\pi$ 
channel through the $\sigma$ meson exchange model of Ref.~\cite{Bernard:1996gq}. 
The resonance saturation hypothesis \cite{Ecker:1988te} then allows us to estimate 
how much of the original LECs is due to rescattering and subtract those amounts 
(see Fig.~\ref{fig:contact}).
%
%

We take the NLO $\chi$EFT $\pi N$ amplitude from Ref.~\cite{Alarcon:2012kn} and perform the
partial-wave projection according to the formulas of Ref.~\cite{Frazer:1960zza}.
The LECs appearing in this amplitude at NLO have been determined
through relativistic $\chi$EFT analysis of $\pi N$ scattering with explicit 
$\Delta$'s \cite{Alarcon:2012kn,Siemens:2016jwj}. Performing the adjustment
as described above, we obtain the values $c_i \; (i = 1, 2, 3)$ 
listed in Table~\ref{Tab:LECs} in Appendix~\ref{app:expressions}.
We use the parameters to evaluate the NLO contribution to $J^0_+(t)$ and estimate 
its uncertainty by varying the values in the determined range. 
Numerical results from this procedure will be shown below.

At N2LO accuracy both the $\pi\pi \rightarrow N\bar N$ PWA and
the pion FF involve loop corrections, and the structure of the $\chi$EFT expressions 
becomes considerably richer. At this order $\pi\pi$ rescattering in the $t$-channel
occurs both in the PWA and in the pion FF, so that both functions 
become complex at $t > 4 M_\pi^2$; one should therefore be able to verify explicitly that 
they have the same phase, and that the phase cancels in the ratio in Eq.~(\ref{J_def}). 
Furthermore, at N2LO $\pi N$ and $\pi \Delta$ $s$-channel intermediate states 
appear in the PWA and contribute to its left-hand cut.
Here we do not pursue a full N2LO calculation of the function $J^0_+(t)$ including loops.
Instead, we estimate the size of the N2LO corrections in a simple way, by using the N2LO 
tree level result and varying the LECs in a meaningful range. To this end we impose the 
unsubtracted dispersion relation for the scalar FF at $t = 0$ (sigma term),
\be
\sigma(0) \; = \; \frac{1}{\pi}\int_{4M_\pi^2}^\infty\! dt'\ \frac{\text{Im} \, \sigma(t')}{t'} ,
\label{Eq:Charge_sum_rule}
\ee
with the integration restricted to the region $t' < 1\, \textrm{GeV}^2$. This relation
fixes the LECs in the N2LO tree level result in terms of $\sigma(0)$. We then generate
an uncertainty band by varying $\sigma(0)$ in the range 45--59 MeV.
The first value was determined in an earlier dispersive analysis of the sigma term \cite{Gasser:1990ce},
while the second was obtained by $\chi$EFT from modern $\pi N$ PWAs and pionic atom data \cite{Alarcon:2011zs}.
Numerical results for $J^0_+(t)$ with these parameters will be shown below.
\subsection{Pion form factor}
%
%
\begin{figure}
\begin{center}
\includegraphics[width=.42\textwidth]{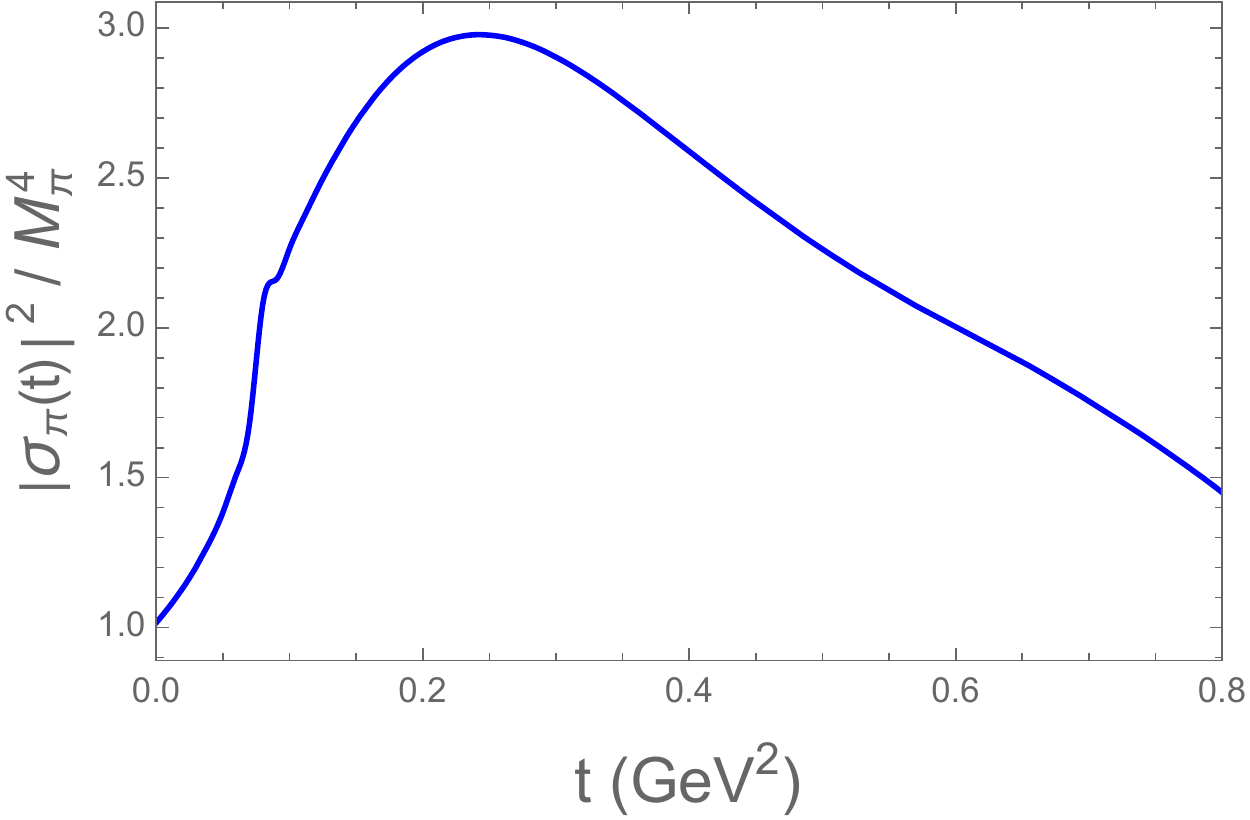}
\caption{\small Pion scalar FF obtained in the dispersive analysis of
Ref.~\cite{Celis:2013xja}. The plot shows the normalized
squared modulus of the FF, $|\sigma_\pi(t)|^2/M_\pi^4$, as it enters
in the improvement formula Eq.~(\ref{improvement}).
\label{fig:pion}}
\end{center}
\end{figure} 
For the pion scalar FF in Eq.~(\ref{improvement}) we take the result of the 
dispersive analysis of Ref.~\cite{Celis:2013xja}; for earlier results see
Refs.~\cite{Ananthanarayan:2004xy,Oller:2007xd}. The analysis includes the 
$K\bar K$ channel at $t > 1\, \textrm{GeV}^2$ in a coupled-channel approach; 
we require only the result on the $\pi\pi$ cut for $t < 1\, \textrm{GeV}^2$.
Figure~\ref{fig:pion} shows the normalized squared modulus of the FF,
$|\sigma_\pi(t)|^2/M_\pi^4$, as it enters in the dispersive improvement 
formula Eq.~(\ref{improvement}). One sees that it reaches a value $\sim 3$ 
at $t \sim 0.3\, \textrm{GeV}^2$, which indicates the presence of strong 
$\pi\pi$ rescattering. This underscores the rationale for our approach, 
as it would be very difficult to incorporate these effects
through higher-order chiral corrections.

The overall uncertainty of the nucleon spectral function calculated in our approach is 
determined by the uncertainty of the $\chi$EFT calculation of $J^0_+(t)$ and the uncertainty 
of the empirical $\sigma_\pi(t)$. In the numerical results presented in the following we 
show only the uncertainties resulting from the $\chi$EFT calculation of $J^0_+(t)$,
which can be quantified within our approach.
\section{Results}
\label{sec:results}
\subsection{Nucleon scalar spectral function}
\begin{figure}
\begin{center}
\epsfig{file= 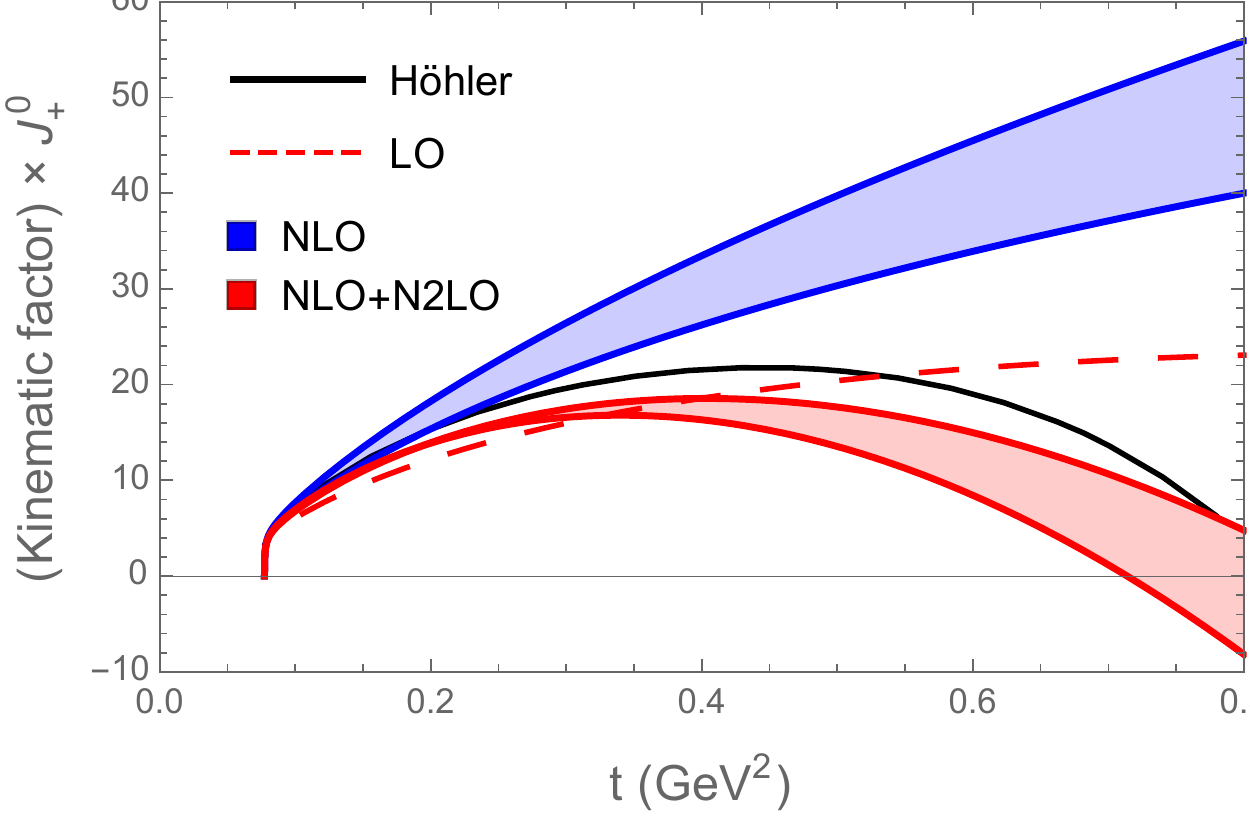,width=.45\textwidth,angle=0}
\caption{\small $\chi$EFT results for the function 
$3 k_{\rm cm}/(4 \widetilde p_N^2 \sqrt{t}) \, J^0_+(t)$, Eq.~(\ref{unitarity_real}).
Dashed line: LO. Red band: NLO. Blue band: NLO + N2LO, estimated as described in Sec.~\ref{subsec:higher-order}. 
(The bands labeled NLO and NLO+N2LO show the total result up to that order and include the LO contribution.)
Solid black line: Dispersion-theoretical result of Ref.~\cite{Hoehler}.
\label{fig:J}}
\end{center}
\end{figure}

We now present the results of the dispersively improved $\chi$EFT calculation of the scalar FF
with the methods described in Sec.~\ref{sec:method}. Figure~\ref{fig:J} shows
the function $J^0_+(t)$, Eq.~(\ref{J_def}), which is the primary object of the 
$\chi$EFT calculation (cf.~Fig.~\ref{fig:ndelta}). One observes: 
\begin{itemize}
\item[(a)] The chiral expansion shows good convergence. Higher-order correction are small at threshold 
and become increasingly important at larger $t$. NLO corrections from the LECs give a strong positive
contribution at $t > 0.5\, \textrm{GeV}^2$ (mainly due to the contribution of $c_3$), which is 
corrected downwards by the N2LO corrections estimated according to Sec.~\ref{subsec:higher-order}.
As a consequence, the NLO+N2LO results are close to the LO over a wide range of $t$.
\item[(b)] The $\chi$EFT predictions agree well with the dispersion-theoretical result of Ref.~\cite{Hoehler},
obtained by analytic continuation of the $\pi\pi \rightarrow N\bar N$ PWA extracted from 
$\pi N$ scattering data. The LO $\chi$EFT result describes the dispersion-theoretical result
very close to threshold. The NLO corrections improve the behavior in the near threshold-region 
and lead to agreement with the dispersion-theoretical result up to $t\lesssim 0.2 \textrm{GeV}^2$, 
but are too large at larger $t$. Finally, the N2LO estimate reproduces the dispersion-theoretical
result over the entire range up to $t\sim 0.8 \textrm{GeV}^2$.
\end{itemize}
These observations provide strong justification for our program of applying $\chi$EFT to the real 
function $J^0_+(t)$, including the approximate treatment of N2LO corrections 
(see Sec.~\ref{subsec:higher-order}). The convergence
pattern observed here directly carries over to the spectral function $\textrm{Im}\, \sigma(t)$.
%
%
%

%
%
\begin{figure}
\begin{center}
\epsfig{file=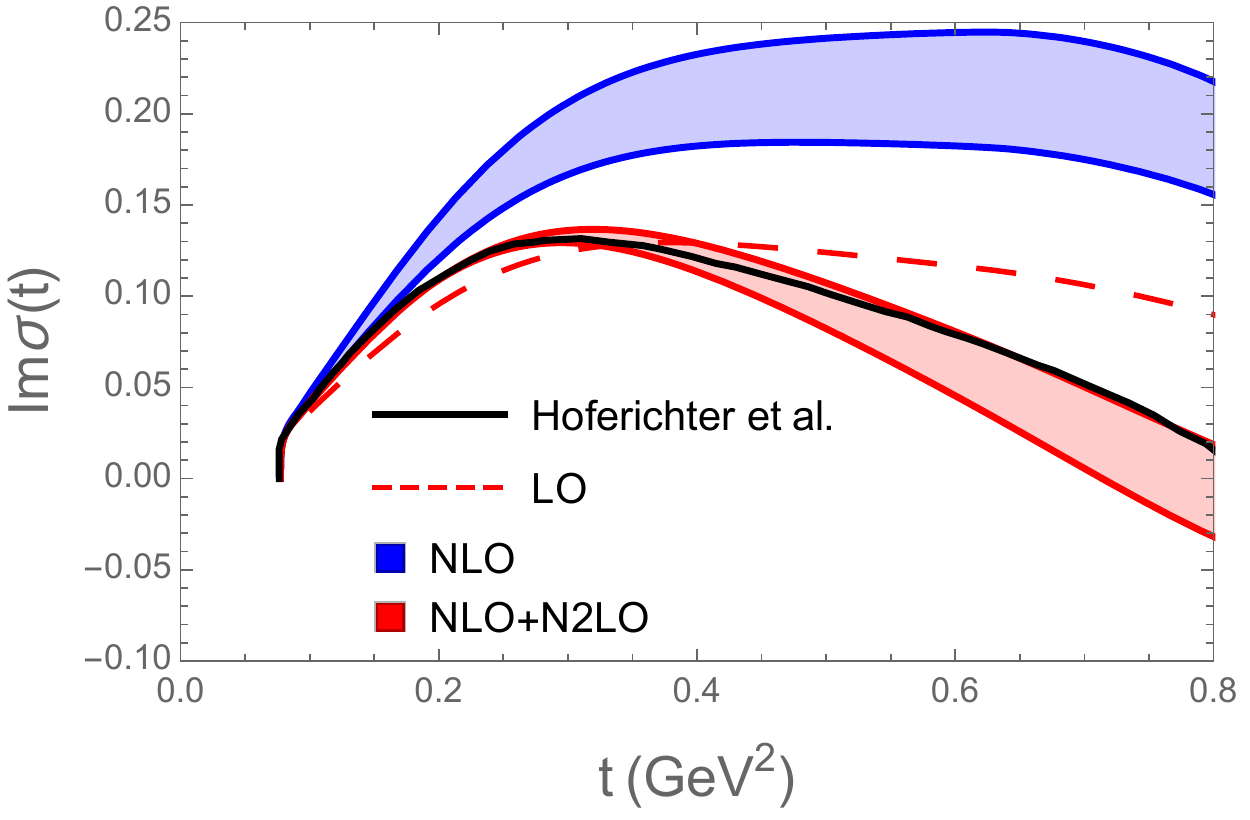,width=.48\textwidth,angle=0}
\epsfig{file=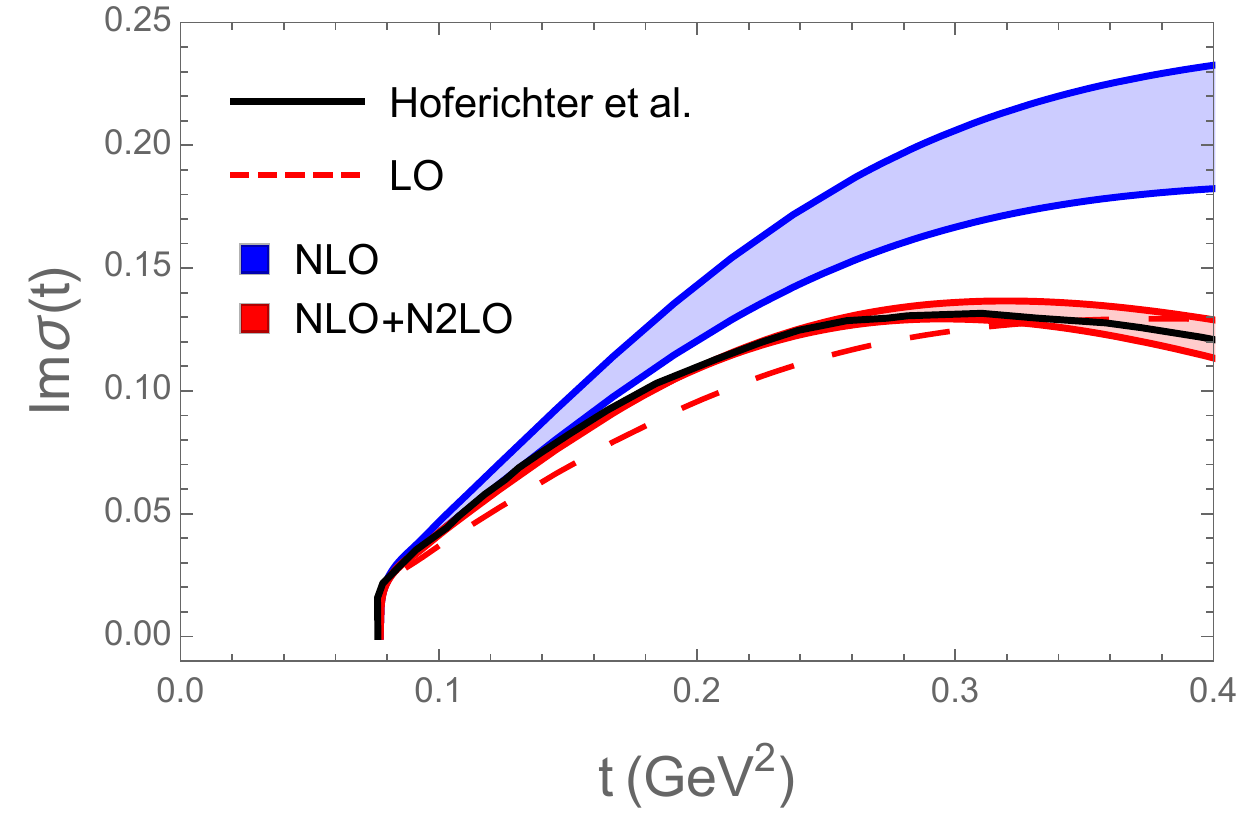,width=.48\textwidth,angle=0}
\caption{\small DI$\chi$EFT results for the scalar spectral function, Eq.~(\ref{improvement}). 
The upper plot covers the full range up to $t = 0.8\, \textrm{GeV}^2$; the lower plot covers
the near-threshold region. The dashed curves and bands 
showing the LO, NLO, and NLO+N2LO approximations correspond to those of Fig.~\ref{fig:J}.
Solid lines: Roy-Steiner result of Ref.~\cite{Hoferichter:2012wf}.\label{Fig:Scalar_spectral_function}}
\end{center}
\end{figure} 
Figure~\ref{Fig:Scalar_spectral_function} shows our predictions for the scalar spectral function 
$\textrm{Im}\,\sigma(t)$, obtained by multiplying the $\chi$EFT results for $J^0_+(t)$
with the empirical $|\sigma_\pi(t)|^2$, Eq.~(\ref{unitarity_real}). Also shown is the spectral 
function obtained from a recent analysis using Roy-Steiner 
equations \cite{Hoferichter:2012wf}.\footnote{We compare our results for $J^0_+(t)$ with
the dispersion-theoretical analysis of Ref.~\cite{Hoehler}, which is based on old data but
quotes results directly for this real function. Our results for $\textrm{Im}\, \sigma(t)$ we rather
compare with the Roy-Steiner analysis of Ref.~\cite{Hoferichter:2012wf}, which is based on the most
recent $\pi\pi$ and $\pi N$ scattering data.} One observes that
the LO $\chi$EFT result is in reasonable agreement with the Roy-Steiner result up to 
energies $t\sim 0.3\, \textrm{GeV}^2$. The NLO correction improves the near-threshold behavior 
but overestimates the spectral function at intermediate energies.
The N2LO corrections, estimated according to Sec.~\ref{subsec:higher-order}, 
have the right $t$-dependence to correct this issue. Altogether we obtain
excellent agreement with the Roy-Steiner result up to $t \sim 1 \, \textrm{GeV}^2$. 
\subsection{Nucleon scalar radius}
With the DI$\chi$EFT result for the spectral function we can now compute the nucleon scalar radius,
using the  well-convergent dispersion integral Eq.~(\ref{r2_dispersive}).
Table~\ref{tab:scalar_radius} shows the results obtained with different values of $\sigma(0)$.
Note that $\sigma(0)$ enters directly in the normalization factor, Eq.~(\ref{r2_dispersive}),
and indirectly through the procedure fixing the N2LO parameters (see Sec.~\ref{subsec:higher-order}).
One observes that the nucleon scalar radius is substantially larger than the charge radius 
(Dirac radius), $\langle r^2\rangle_\sigma > \langle r^2 \rangle_1
\sim 0.65 \, \textrm{fm}^2$, as pointed out in Ref.~\cite{Gasser:1990ap}.

It is interesting to compare our results for the scalar radius with those of the 
dispersion-theoretical calculation of Ref.~\cite{Gasser:1990ap}, not the least
because the DI$\chi$EFT calculation can provide systematic uncertainty estimates.
We find that, at NLO+N2LO accuracy, our radius calculated with $\sigma(0)$ = 45 MeV 
is smaller than that of Ref.~\cite{Gasser:1990ap}, which uses the same value of $\sigma(0)$.
The difference can be traced back to the spectral function, which in our calculation 
comes out smaller than that of Ref.~\cite{Gasser:1990ap} in the near-threshold region.
We note that the DI$\chi$EFT result agrees with that of the Roy-Steiner analysis of 
Ref.~\cite{Hoferichter:2012wf}.
%
%
\begin{table}
\begin{ruledtabular}
\begin{tabular}{rcccc}
                                             &    &LO & NLO & NLO+N2LO\\
$\langle r^2 \rangle_\sigma$ (fm$^2$) 
& A & 1.06 & 1.40--1.67 & 1.03--1.13\\
& B & 1.38 & 1.83--2.19 & 1.34--1.49\\
\end{tabular}
\caption{Nucleon scalar radius obtained in DI$\chi$EFT with different values of 
$\sigma(0)$. A) $\sigma(0) = 59$~MeV; B) 45~MeV. 
\label{tab:scalar_radius}}
\end{ruledtabular}
\end{table}
\subsection{Nucleon scalar form factor}
\begin{figure}
\begin{center}
 \epsfig{file=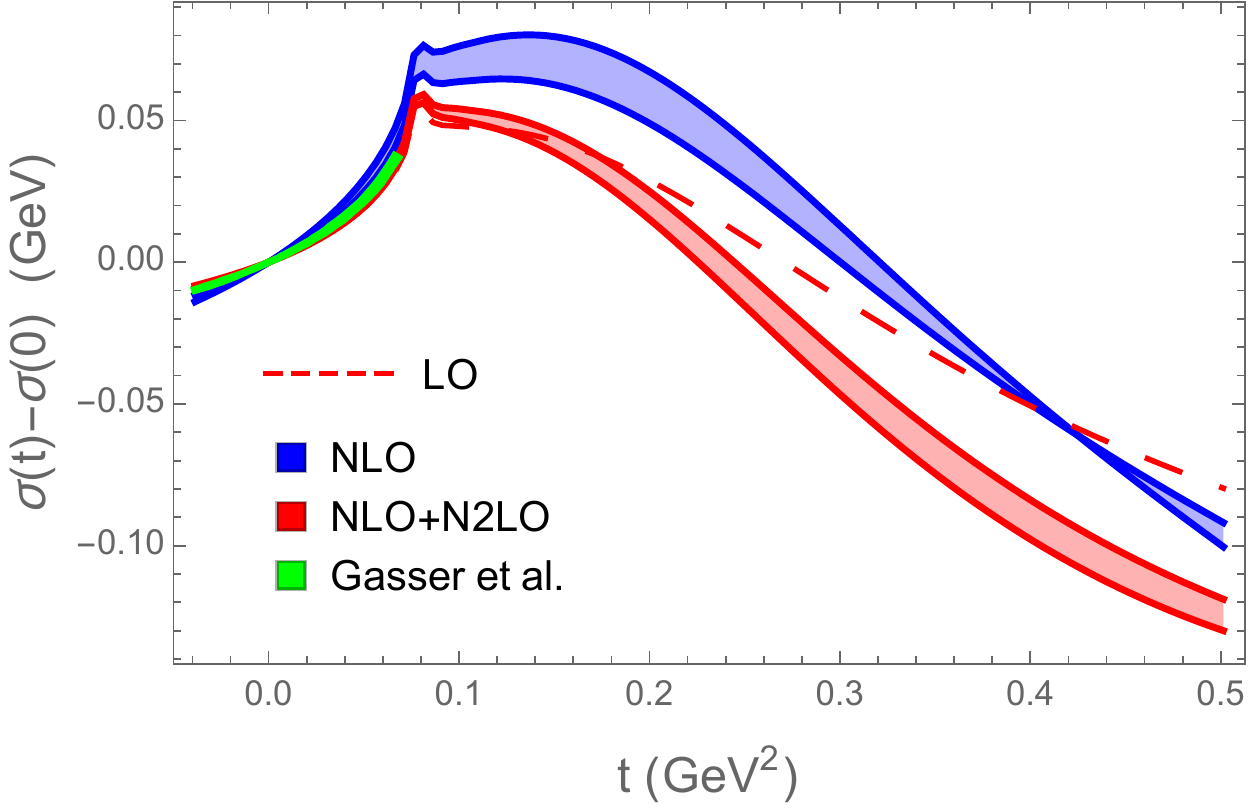,width=.45\textwidth,angle=0} 
 \epsfig{file=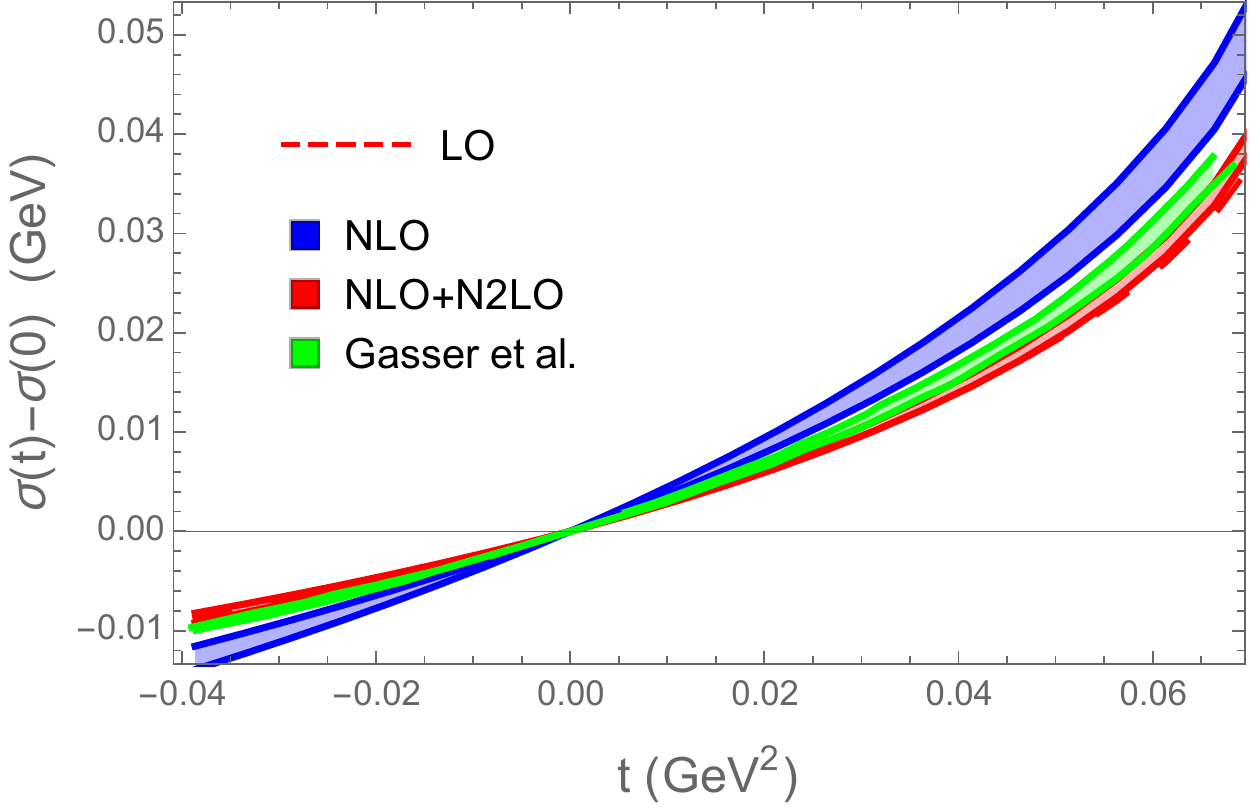,width=.45\textwidth,angle=0} 
\caption{\small Nucleon scalar FF $\Delta \sigma(t) = \sigma(t) - \sigma(0)$,
obtained from the once-subtracted dispersion integral Eq.~(\ref{ff_dispersive}) with the
DI$\chi$EFT spectral functions. The dashed lines, blue bands, and red bands show the LO, NLO and NLO+N2LO
approximations (cf.\ Fig.~\ref{Fig:Scalar_spectral_function}).
The green bands show the dispersion-theoretical result of Ref.~\cite{Gasser:1990ap}.  
\label{Fig:Scalar_form_factor}}
\end{center}
\end{figure}

Using the once-subtracted dispersion relation Eq.~(\ref{ff_dispersive}) we can also calculate 
the $t$-dependent scalar FF, both in the region below threshold $t < 4 M_\pi^2$ (where it is real) 
and above threshold $t > 4 M_\pi^2$ (where it is complex). Figure~\ref{Fig:Scalar_form_factor} 
shows the results obtained with the DI$\chi$EFT spectral functions at different orders,
along with the dispersion-theoretical result of Ref.~\cite{Gasser:1990ap}. In order to suppress 
the dependence on the uncertain $\sigma(0)$ the figure shows the difference $\sigma(t)-\sigma(0)$ 
instead of $\sigma(t)$.
One observes that the DI$\chi$EFT calculations converge well, especially at $t < 4 M_\pi^2$. 
The LO approximation already gives a result in good agreement with the dispersive one. 
The NLO contribution corrects the LO result in the right direction, but by too much in magnitude;
this is because it overestimates the spectral function in the intermediate-$t$ region, which still 
has some modest influence on the result of the dispersion integral for the FF. 
At NLO+N2LO, once we enforce that the dispersion integral reproduce the chosen $\sigma (0)$
(see Sec.~\ref{subsec:higher-order}), the DI$\chi$EFT scalar FF is in excellent 
agreement with the dispersion-theoretical one.
%
%
\subsection{Cheng-Dashen discrepancy}
Table~\ref{Tab:Delta_sigma} gives the results for the Cheng-Dashen discrepancy Eq.~(\ref{Delta}) 
in DI$\chi$EFT, calculated through the dispersion integral Eq.(\ref{Delta_dispersive}).
One observes the same pattern of convergence as in the scalar radius and the FF: The NLO 
corrections are strongly positive, the N2LO corrections are negative, such that the total
result at N2LO is rather close to the original LO one.

The Cheng-Dashen discrepancy has been computed previously using different methods. 
The first $\chi$EFT calculation was reported in Ref.~\cite{Gasser:1987rb} and obtained 
$\Delta_\sigma = 4.6~\text{MeV}$ at $\mathcal{O}(p^3)$ accuracy. The dispersive analysis 
of Ref.~\cite{Gasser:1990ap} then obtained a much larger value, $\Delta_\sigma = 15.2(4)$~MeV, 
pointing to the inability of $\chi$EFT to generate sufficient curvature in $\sigma(t)$ 
at $\mathcal{O}(p^3)$ accuracy (according to the the same reference, 
this has almost no effect on the extraction of the sigma term from $\pi N$ scattering data).
The curvature necessary to recover the dispersive result was obtained in an $\mathcal{O}(p^4)$ 
calculation in Ref.~\cite{Becher:2001hv}, which found $\Delta_\sigma = 14.0~\text{MeV}+2 M_\pi^4 \bar{e}_2$.
This larger value was supported by an updated dispersive calculation in Ref.~\cite{Hoferichter:2012wf}, 
which finds $\Delta_\sigma = 13.9(3)$~MeV. The DI$\chi$EFT approach described here gives results 
in excellent agreement with the dispersive calculation. The main improvement compared to conventional 
$\chi$EFT is that it includes the strong $\pi\pi$ rescattering effects. It is interesting that
such effects are essential even at $t=2M_\pi^2$, which should be well within the radius of 
convergence of conventional $\chi$EFT calculations.
\begin{table}
\begin{ruledtabular}
\begin{tabular}{cccc}
                                                &LO & NLO & NLO+N2LO\\
  $\Delta_\sigma$ (MeV)& 13.3 & 17.4 - 20.6 & 13.3 - 14.5 \\
\end{tabular}
\caption{DI$\chi$EFT results for the Cheng-Dashen discrepancy $\Delta_\sigma$, Eq.~(\ref{Delta}). 
\label{Tab:Delta_sigma}}
\end{ruledtabular}
\end{table}
\section{Discussion}
\label{sec:discussion}
\subsection{Euclidean correlation functions}
\label{subsec:euclidean}
The DI$\chi$EFT approach incorporates $\pi\pi$ rescattering effects in the nucleon spectral 
functions through the timelike pion FF, which is provided by sources 
outside of $\chi$EFT. An important aspect is that the pion FF enters only through 
its squared modulus, and that knowledge of its phase is not required [see Eq.~(\ref{unitarity_real})].
The modulus of the pion timelike FF can in principle be extracted from the vacuum correlation function 
of the operator, which can be continued to imaginary time (Euclidean QCD) and evaluated using 
non-perturbative methods such as Lattice QCD. This opens up the interesting possibility of 
combining the $\chi$EFT calculations of the $\pi\pi\rightarrow N\bar N$ amplitude with 
Euclidean QCD calculations of the pion timelike FF. Here we describe this connection for
the scalar operator; the expressions can easily be generalized to other $G$-parity-even operators.

\begin{figure}[t]
\begin{center}
\epsfig{file=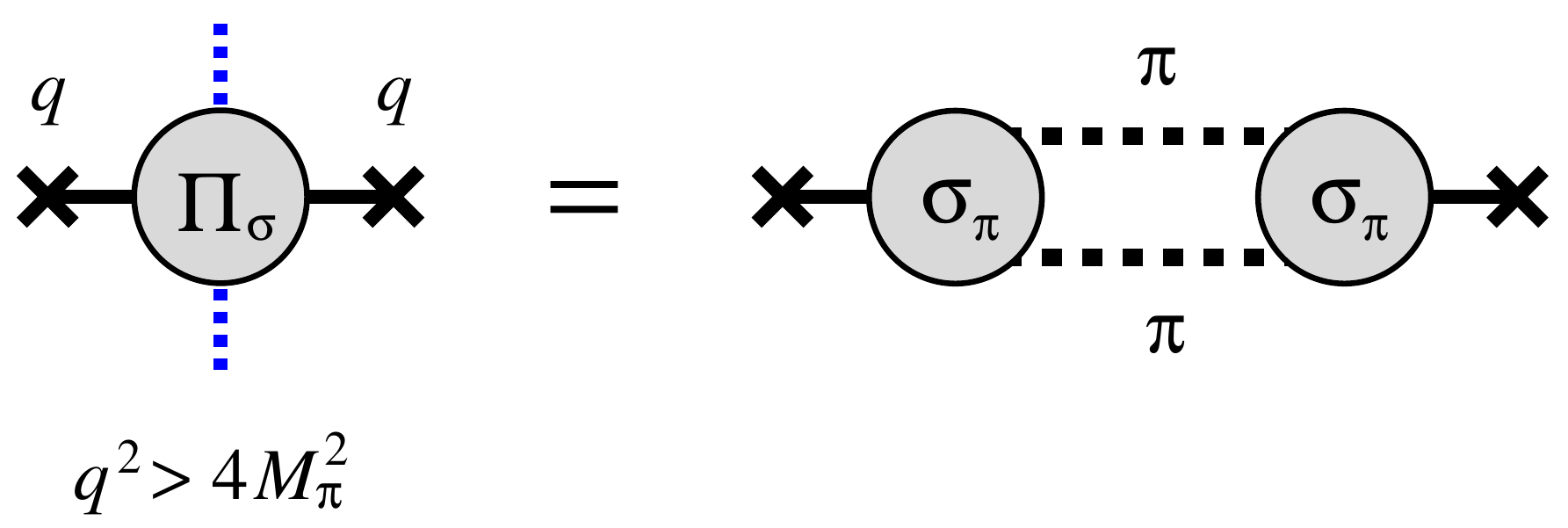,width=.4\textwidth,angle=0} 
\caption{\small Two-pion cut of the scalar correlation function Eq.~(\ref{corr_imaginary_scalar}).}
\label{fig:corr_scalar}
\end{center}
\end{figure}

The vacuum polarization induced by the scalar density operator Eq.~(\ref{operator_def})
is given by the two-point correlation function \cite{Shifman:1978bx,Shifman:1978by,Reinders:1984sr} 
\be
\Pi_\sigma(q^2) \; \equiv \; i \int d^4 x \; e^{i q x} \;
\langle 0| \, \textrm{T} \, O_\sigma(x) O_\sigma(0) \, | 0 \rangle ,
\label{corr}
\ee
where $x$ is a 4-dimensional Minkowskian space-time displacement, T denotes the time-ordering 
operation, and the 4-momentum $q$ can be spacelike or timelike, $q^2 < 0$ or $> 0$. $\Pi_\sigma(q^2)$
is an analytic function of $q^2$, with no singularities at $q^2 < 0$ and cuts at $q^2 > 0$, 
corresponding to hadronic intermediate states produced by the operator $O_\sigma$. The function
obeys subtracted dispersion relations of the form [$\Pi_\sigma^{(k)}(q^2)$ denotes the $k$'th 
derivative with respect to $q^2$]
\ba
&& \Pi_\sigma(q^2) \; - \; \sum\limits_{k = 0}^{n-1} (q^2)^k \; \frac{\Pi_\sigma^{(k)}(0)}{k!}
\nonumber \\
&=& \frac{(q^2)^n}{\pi} \int_{4 M_\pi^2}^\infty 
dt' \; \frac{\textrm{Im} \, \Pi_\sigma (t')}{(t')^n (t' - q^2)} ,
\label{corr_dispersion}
\ea
where $n \geq 2$ for the scalar operator, based on the expected short-distance behavior of the
coordinate-space correlation function. The imaginary part is proportional to the cross section for hadron 
production by the scalar operator at the squared mass $t'$ and is positive, 
$\textrm{Im} \, \Pi_\sigma (t') > 0$. In the region $4 M_\pi^2 < t' < 16 M_\pi^2$ the only 
accessible hadronic state is the $\pi\pi$ state (two-pion cut). The imaginary part on the
two-pion cut is given by an elastic unitarity formula analogous to Eq.~(\ref{unitarity}),
with the pion FF appearing both in the initial (operator $\rightarrow \pi\pi$)
and the final ($\pi\pi \rightarrow$ operator) amplitudes (see Fig.~\ref{fig:corr_scalar}),
\ba
\textrm{Im} \, \Pi_\sigma (t') &=& \frac{k_{\rm cm}}{8\pi \sqrt{t}} \; |\sigma_\pi(t')|^2 
\nonumber \\[1ex]
&& (4 M_\pi^2 < t' < 16 M_\pi^2).
\label{corr_imaginary_scalar}
\ea
It provides a direct connection between the squared modulus $|\sigma_\pi(t')|^2$ and the
vacuum correlation function.
%
%

In order to put Eq.~(\ref{corr_imaginary_scalar}) to practical use one must have a method
to extract the imaginary part on the two-pion cut from (approximate) calculations of the
correlation function Eq.~(\ref{corr}). Here one may use the fact that the lowest-mass state
in the spectral representation determines the asymptotic behavior of the coordinate-space
correlation function at large spacelike distances. Substituting the spectral representation
Eq.~(\ref{corr_dispersion}) of $\Pi_\sigma (q^2)$ in in Eq.~(\ref{corr}) and inverting 
the Fourier transform, one obtains a spectral representation of the coordinate-space function
at spacelike distances,\footnote{In deriving Eq.~(\ref{corr_position_spectral}) one may
disregard the subtractions in Eq.~(\ref{corr_dispersion}) and consider the Fourier transform 
of the formal unsubtracted dispersion integral with $n = 0$. The subtraction terms result 
in delta functions at $x = 0$, or derivatives thereof, which can be neglected
when considering the behavior of the coordinate-space correlation function
at finite distances.}
\ba
&& \langle 0| \, \textrm{T} \, O_\sigma(x) O_\sigma(0) \, | 0 \rangle 
\nonumber \\[1ex]
&=& \int_{4 M_\pi^2}^\infty dt' \; \frac{\sqrt{t'} \, K_1(\sqrt{t'}\sqrt{-x^2})}
{4\pi^2 \sqrt{-x^2}} 
\; \textrm{Im} \, \Pi_\sigma (t') 
\nonumber 
\\[1ex]
&& (x^2 < 0).
\label{corr_position_spectral}
\ea
The modified Bessel function decays exponentially at large arguments,
$K_1 (z) \sim [\pi/(2 \sqrt{z})]^{1/2} \, \exp(-z)$ for $z \gg 1$.
For a given distance $\sqrt{-x^2}$ the factor $K_1(\sqrt{t'}\sqrt{-x^2})$
strongly suppresses the contribution from energies $\sqrt{t'} \gg 1/\sqrt{-x^2}$ in 
Eq.~(\ref{corr_position_spectral}). The asymptotic behavior of the coordinate-space
function is therefore dominated by $t'$ in the vicinity of the two-pion threshold in the spectral
integral; it is of the form $\sim \exp(-2 M_\pi \sqrt{-x^2}) \; P(-x^2)$, where the
pre-exponential factor $P$ depends on the threshold behavior of $\textrm{Im}\, \Pi_\sigma(t')$,
Eq.~(\ref{corr_imaginary_scalar}). 
At large but finite distances the spectral integral Eq.~(\ref{corr_position_spectral})
extends over the entire two-pion cut, with exponential suppression of higher-mass states. 
Whether this representation could be used to extract quantitative information on 
$|\sigma_\pi(t')|^2$ from Lattice QCD calculations of the coordinate-space correlation function 
at distances $\sqrt{-x^2} \sim 1/M_\pi$
is an interesting question for further study. The success of this program depends on the 
contribution of higher-mass states with $t' > 16\, M_\pi^2$ to the spectral integral, 
which could be inferred from the Lattice QCD calculations of the correlator at 
shorter distances, or estimated using quark-hadron duality.

The timelike pion FF at $4 M_\pi^2 < t < 16 M_\pi^2$ can also be computed in Lattice QCD
using a variant of the L\"uscher method, which exploits the correspondence between the
$\pi\pi$ scattering phase shift and the energy levels of the $\pi\pi$ system in a finite 
volume \cite{Meyer:2011um}.
Again this method delivers the squared modulus of the timelike pion FF without determining 
the phase. Results for $|\sigma_\pi(t')|^2$ obtained with either of the methods described here
could be incorporated into our DI$\chi$EFT approach through Eq.~(\ref{unitarity_real}).
\subsection{Connection with large-$N_c$ QCD}
\label{subsec:largenc}
It is worthwhile to investigate the connection of our approach with the large-$N_c$ limit of QCD.
This exercise shows that the DI$\chi$EFT results obey the general $N_c$-scaling relations for the
scalar FF and explains the relative contribution of $N$ and $\Delta$ intermediate states 
in the chiral processes.

The limit of a large number of colors is a powerful method for connecting properties of mesons and 
baryons with the microscopic theory of strong interactions \cite{tHooft:1973alw,Witten:1979kh,Coleman:1980mx}; 
see Ref.~\cite{Jenkins:1998wy} for a review. While the dynamics remains complex and cannot be solved exactly, 
the scaling behavior of meson and baryon properties with $N_c$ can be established on general grounds 
and provides insights into their structure and guidance for the formulation of effective theories.
The masses of low-lying mesons scale as $\mathcal{O}(N_c^0)$; the masses of baryons scale as 
$\mathcal{O}(N_c)$ for states with spin/isospin $\mathcal{O}(N_c^0)$; while the hadronic size 
of mesons and baryons is $\mathcal{O}(N_c^0)$ and remains 
stable in the large-$N_c$ limit. Baryons thus are heavy objects of finite size, whose external motion
in coordinate and isospin/spin space can be described classically, with a mass and moment of inertia 
of $\mathcal{O}(N_c)$. The $N$ and $\Delta$ are the rotational states of the classical body with isospin/spin 
$I = J = 1/2$ and $3/2$, and the mass splitting is $m_\Delta - m_N = \mathcal{O}(N_c^{-1})$. Further scaling
relations can be obtained for the matrix elements of QCD operators between meson and baryon states,
and the meson-meson and meson-baryon couplings. The relations are model-independent and can be 
derived in many different ways: diagrammatic arguments \cite{Witten:1979kh}, group-theoretical 
methods \cite{Dashen:1993jt,Dashen:1994qi}, large-$N_c$ quark models \cite{Karl:1984cz,Jackson:1985bn}, 
and the soliton picture of baryons \cite{Adkins:1983ya,Zahed:1986qz}. 

The $N_c$-scaling of the nucleon's scalar FF considered in the present study can be established as
follows. The standard techniques show that the nucleon sigma term scales as $\sigma(0) = \mathcal{O}(N_c)$.
This is plausible because the sigma term represents the response of the nucleon mass to a change of the 
QCD quark mass, $\sigma (0) = \hat m (\partial / \partial \hat m) m_N$. Since the nucleon's spatial 
size is $\mathcal{O}(N_c^0)$, we consider the scalar FF at non-exceptional momentum transfers 
$|t| = \mathcal{O}(N_c^0)$, in either the spacelike or timelike domain. For such values of 
$t$ the scaling behavior of the FF is then given by
\be
\sigma(t) \;  = \; \mathcal{O}(N_c) \hspace{2em} [|t| = \mathcal{O}(N_c^0)].
\label{sigma_largen}
\ee
Because $M_\pi = \mathcal{O}(N_c^0)$, the $t$-region of elastic unitarity ($4 M_\pi^2 < t < 16 M_\pi^2$)
remains stable in the large-$N_c$ limit. Thus the basic setup of our dispersive analysis remains stable 
in the large-$N_c$ limit: the dispersion integral extends over momenta $t = \mathcal{O}(N_c^0)$ and converges
in that parametric domain, and we require the spectral function at energies $t = \mathcal{O}(N_c^0)$.
For the pion scalar FF the same arguments lead to $\sigma_\pi (0) = M_\pi^2 = \mathcal{O}(N_c^0)$
and therefore
\be
\sigma_\pi (t) \;  = \; \mathcal{O}(N_c^0) \hspace{2em} [|t| = \mathcal{O}(N_c^0)].
\label{sigma_largen_pion}
\ee
Equations~(\ref{sigma_largen}) and (\ref{sigma_largen_pion}) imply that the scalar 
FF of the hadrons scales with the number of their valence quarks, as one would
expect in a ``constituent quark'' picture of dynamical chiral symmetry breaking.

It is easy to verify the that the DI$\chi$EFT results for the scalar nucleon FF obey the
general $N_c$-scaling of Eq.~(\ref{sigma_largen}). Using the explicit expression
Eq.~(\ref{J_LO_N}) with $g_A = \mathcal{O}(N_c)$ and $f_\pi = \mathcal{O}(\sqrt{N_c})$, one finds that 
the contribution of the $N$ Born term to $J_+^0(t)$ scales as
\be
J_+^0(t)[\textrm{LO}, N] \; =\;  \mathcal{O}(N_c^3) \hspace{2em} [|t| = \mathcal{O}(N_c^0)].
\label{J_largen_eft}
\ee
Using Eq.~(\ref{J_LO_Delta}) with $h_A = \mathcal{O}(N_c)$ one finds the same scaling behavior for 
the contribution of the $\Delta$ Born term.\footnote{Equation~(\ref{J_largen_eft}) 
is valid for $k_{\rm cm}^2 = t/4 - M_\pi^2 = \mathcal{O}(N_c^0)$, i.e., for non-exceptional 
values $t = \mathcal{O}(N_c^0)$, not parametrically close to the two-pion threshold.
In this case $\{ x_N, x_\Delta \} = \mathcal{O}(N_c^{-1})$ in Eqs.~(\ref{x_N}) and (\ref{x_Delta}), 
and the inverse tangent functions in Eqs.~(\ref{J_LO_N}) and (\ref{J_LO_Delta}) 
count as $\arctan \, \{ x_N, x_\Delta \} \approx \pi/2 = \mathcal{O}(N_c^0)$. The polynomial
terms in Eqs.~(\ref{J_LO_N}) and (\ref{J_LO_Delta}) are suppressed relative to
the inverse tangent terms by a power $N_c^{-1}$.} 
Multiplication with $|\sigma_\pi(t)|^2$, Eq.~(\ref{sigma_largen_pion}),
does not change this scaling behavior. Taking into account the scaling behavior of the kinematic 
factor in Eq.~(\ref{unitarity}) one concludes that the DI$\chi$EFT result 
for the spectral function scales according to Eq.~(\ref{sigma_largen}). This scaling behavior
then carries over to the FF through the dispersion relation Eq.~(\ref{ff_dispersive}).

It is interesting to compare the relative contributions of the $N$ and $\Delta$ Born terms
to the scalar spectral function in the large-$N_c$ limit. In the large-$N_c$ limit the 
$N$ and $\Delta$ become degenerate, $\{ m_N, m_\Delta \} = \mathcal{O}(N_c)$ and 
$m_\Delta - m_N = \mathcal{O}(N_c^{-1})$, and the $\pi NN$ and $\pi N\Delta$ couplings 
are related by \cite{Adkins:1983ya}
\be
g_{\pi N\Delta} \; = \; \frac{3}{2} \; g_{\pi NN} \hspace{2em} (N_c \rightarrow \infty ) .
\label{g_pindelta_nc}
\ee
Both statements follow from the fact that the $N$ and $\Delta$ are rotational states 
of a classical body with combined isospin/spin symmetry. The conventional couplings of 
Eq.~(\ref{g_pindelta_nc}) are related to the $\chi$EFT couplings by
\be
g_{\pi NN} \; = \; \frac{g_A m_N}{f_\pi},
\hspace{2em} \; g_{\pi N \Delta} \; = \; \frac{h_A m_N}{\sqrt{2}f_\pi}.
\label{g_eft}
\ee
and scale as $\{ g_{\pi NN}, g_{\pi N\Delta} \} = \mathcal{O}(N_c^{3/2})$. 
Using Eqs.~(\ref{g_pindelta_nc}) and (\ref{g_eft}) and the expressions in 
Appendix~\ref{app:expressions} one easily shows that the LO $\chi$EFT results satisfy
\be
\textrm{Im}\, \sigma(t) [\textrm{LO}, \Delta] \; = \; 
2 \, \textrm{Im}\, \sigma(t) [\textrm{LO}, N] 
\hspace{1.5em} (N_c \rightarrow \infty ) ,
\label{scalar_largen_n_delta}
\ee
i.e., the contribution of the $\Delta$ Born term is twice as large as that of the $N$ one. 
Such behavior was observed in earlier $\chi$EFT calculations of the nucleon's scalar structure
\cite{Cohen:1992uy,Cohen:1996zz}. It was also seen in studies of the nucleon's peripheral 
gluon and singlet quark structure, which are measured by operators with the same isospin-spin 
quantum numbers as the scalar density \cite{Strikman:2003gz,Strikman:2009bd}.

The relative factor in Eq.~(\ref{scalar_largen_n_delta}) can be explained in a simple manner.
Consider the scalar FF in the proton isospin state, $p$. In the Born graph with intermediate $N$
the possible intermediate states are $\pi^0 p$ and $\pi^+ n$; in the Born graph with intermediate 
$\Delta$ they are $\Delta^{++} \pi^-, \Delta^+ \pi^0$, and $\Delta^0 \pi^+$. The Lagrangian with
the relevant $\pi NN$ and $\pi N\Delta$ couplings is
\ba
\mathcal{L}_{\pi N B} &\propto& \frac{g_{\pi NN}}{2} (\sqrt{2} \bar p n \pi^+  + \bar p p \pi^0 )
\nonumber
\\[1ex]
&+& \frac{g_{\pi N\Delta}}{3} (\sqrt{3} \bar p \Delta^{++} \pi^-  + \sqrt{2} \bar p \Delta^0 \pi^0  
+ \bar p \Delta^0 \pi^+ )
\nonumber
\\[1ex]
&+& \textrm{(hermitean conjugate),}
\label{lagrangian_n_delta}
\ea
where we display only the isospin structure and omit the dependence on the pion momentum (for the
full structure see Ref.~\cite{Granados:2016jjl} and references therein). Each pion state 
contributes equally to the scalar charge, cf.~Eq.~(\ref{scalar_pion}). The relative contribution
of the $N$ and $\Delta$ Born graphs is therefore given by the sum of the squared couplings
in Eq.~(\ref{lagrangian_n_delta}),
\ba
\textrm{Im}\, \sigma(t) [\textrm{LO}, N] &\propto & 
\frac{g_{\pi NN}^2}{4}(2 + 1) = \frac{3 g_{\pi NN}^2}{4} ,
\\
\textrm{Im}\, \sigma(t) [\textrm{LO}, \Delta] &\propto & 
\frac{g_{\pi N\Delta}^2}{9}(3 + 2 + 1) = \frac{2 g_{\pi N\Delta}^2}{3} .
\ea
With the large-$N_c$ relation between the couplings, Eq.~(\ref{g_pindelta_nc}), one
then obtains the relative factor of Eq.~(\ref{scalar_largen_n_delta}).

The observation of Eq.~(\ref{scalar_largen_n_delta}) represents one instance of a general
phenomenon in $\chi$EFT: In the large-$N_c$ limit the contributions of $N$ and $\Delta$
intermediate states are related by a simple factor, and their sum produces a result that
exhibits correct $N_c$-scaling established on general grounds. Note that the factor 
between the $N$ and $\Delta$ contributions depends on the isospin-spin quantum numbers of 
the operator and the matrix element \cite{Cohen:1992uy,Cohen:1996zz}. 
For the scalar operator considered here, the
individual $N$ and $\Delta$ contributions already have the correct $N_c$ scaling,
and summing them just increases the coefficient compared to $N$ only. In the case
of the isovector-vector FFs (Dirac and Pauli), the individual $N$ and $\Delta$ contributions 
have incorrect $N_c$-scaling --- their scaling exponents are too large by one power of $N_c$ --- 
and summing them is necessary in order to cancel the leading term and recover the correct
general $N_c$-scaling Refs.~\cite{Cohen:1992uy,Cohen:1996zz,Granados:2016jjl,Granados:2013moa}.

The large-$N_c$ relation Eq.~(\ref{scalar_largen_n_delta}) represents an important theoretical 
constraint on our $\chi$EFT calculation of scalar nucleon structure. Confronting the large-$N_c$
prediction with the actual $\chi$EFT results for $J^0_+(t)$ obtained with the physical $N$ and $\Delta$ 
masses and couplings, Fig.~\ref{fig:ndelta}, we observe: (a) The actual $N$ and $\Delta$ contributions
have the same sign, in agreement with the large-$N_c$ predictions. (b) The magnitude of the actual
$\Delta$ contribution is significantly smaller than the large-$N_c$ prediction, amounting to $\sim 1/2$
rather than $2$ times the $N$ contribution. This demonstrates that $1/N_c$ suppressed terms play
an essential role in the $\chi$EFT result for $J^0_+(t)$. Notice that the large-$N_c$ limit 
corresponds to the heavy-baryon limit of $\chi$EFT because $M_\pi / m_N = \mathcal{O}(N_c^{-1})$;
it is known that the heavy-baryon expansion converges poorly for the spectral functions
on the two-pion cut; see Refs.~\cite{Becher:1999he,Granados:2013moa} for a discussion. 
\section{Outlook}
\label{sec:outlook}
We have presented a general method for calculating the nucleon FFs of $G$-parity-even operators 
by combining $\chi$EFT and dispersion theory. The spectral functions on the two-pion cut are
constructed with the help of the elastic unitarity condition, using a manifestly real representation 
that separates the coupling of the $\pi\pi$ system to the nucleon from the $\pi\pi$ rescattering
($N/D$ method). $\chi$EFT is used to calculate the real function describing the $\pi\pi$ coupling 
to the nucleon, which is free of $\pi\pi$ rescattering effects. It is dominated by the LO Born
amplitudes with $N$ and $\Delta$ intermediate states and shows good convergence in higher orders.
The effects of $\pi\pi$ rescattering are then incorporated by multiplying with the squared modulus 
of the timelike pion FF, which can be determined empirically or extracted from Lattice QCD calculations 
of the vacuum correlation function of the operator. Our method represents a major improvement over
traditional $\chi$EFT calculations of the spectral functions, which try to account for the $\pi\pi$ 
rescattering effects through $\chi$EFT interactions. It permits calculations of nucleon spectral 
functions up to $t \sim 1\, \textrm{GeV}^2$ (details depend on the operator) and opens up the
prospect of a realistic dispersive analysis of nucleon FFs and related quantities based on $\chi$EFT.

We have applied the method to the nucleon scalar FF. The $\chi$EFT calculations of the real function
$J_+^0(t)$ show good convergence and are in excellent agreement with dispersion-theoretical results
up to $t \sim 0.8\, \textrm{GeV}^2$. This information is sufficient for evaluating the $t$-dependence
of the scalar FF, the scalar radius, and the Cheng-Dashen discrepancy, through a once-subtracted 
dispersion relation. Our calculation determines the scalar FF at momentum transfers up to 
$|t| \sim 0.5\, \textrm{GeV}^2$ with controlled uncertainties. The nucleon's scalar FF is of principal 
interest for understanding the role of dynamical chiral symmetry breaking in nucleon structure and 
the origin of the nucleon mass. It is also an ingredient in modeling the interaction of dark matter
with the nucleon for the purpose of designing direct detection experiments \cite{Bishara:2016hek}.

The method described here can be applied to nucleon FFs of any $G$-parity-even operators with a 
two-pion cut. Applications to the nucleon isovector-vector FFs will be presented in a forthcoming 
article \cite{Alarcon:vector}. Other possible applications are the nucleon FFs of the energy-momentum 
tensor and the moments of generalized parton distributions; see Ref.~\cite{Pasquini:2014vua} for 
a recent dispersive calculation. The impact of the method depends on the convergence of the
$\chi$EFT calculations of the $J$-functions, and on the actual strength distribution in the dispersion 
integrals under study, and has to be demonstrated channel-by-channel. Subtractions can make the
dispersion integrals more convergent and emphasize the low-$t'$ region where the spectral functions
can be computed using our method. Another attractive possibility is to consider the transverse spatial 
densities associated with the FFs, which are represented by exponentially convergent dispersion 
integrals and can safely be calculated with our method at peripheral distances $b \gtrsim 1 \, M_\pi^{-1}$
\cite{Alarcon:2017asr}. An interesting question is whether the dispersive method described
here could be extended to calculate the nucleon's peripheral partonic structure at fixed light-front 
momentum fraction $x$; such calculations have so far been performed in the standard $\chi$EFT approach
without explicit treatment of $\pi\pi$ rescattering \cite{Strikman:2003gz,Strikman:2009bd,Granados:2015rra}.

An obvious extension of the present calculation would be to the nucleon FFs of the scalar strange 
quark and gluonic operators, which have the same quantum numbers as the light-quark scalar
operator Eq.~(\ref{operator_def}) \cite{Shifman:1978zn}. The dispersive calculation of these 
FFs must include also the $K\bar K$ channel and its coupling to $\pi\pi$ in a coupled-channel approach. 
The extension of our method to this situation raises several interesting questions: (a) One
would need to generalize the $N/D$ method and the manifestly real representation of the unitarity 
condition, Eq.~(\ref{unitarity_real}), to the case of coupled $\pi\pi$ and $K\bar K$ channels,
and possibly other inelasticities. (b) One would need to explore how well $\chi$EFT works
for the coupling of the $K\bar K$ system to the nucleon (octet and decuplet baryon Born terms,
contact terms). (c) The distribution of strength in the dispersive integral of the strange
and gluonic FF is expected to be very different from that of the light-quark scalar FF and
may involve large contributions from energies $t' > 1\, \textrm{GeV}$, where our approach is
not applicable. (d) One would need a parametrization of the pion and kaon FFs of these operators 
that takes into account coupled-channel dynamics. Some experimental information on these FFs is 
available from $\tau$ lepton decays \cite{Celis:2013xja}. The timelike pion and kaon FFs could also
be extracted from the vacuum correlation function of the respective operators, as described
in Sec.~\ref{subsec:euclidean}.

The DI$\chi$EFT approach described here could in principle also be extended to the nucleon FFs
of $G$-parity odd operators with a 3-pion cut. Methods for implementing elastic unitarity 
in 3-body channels are presently being developed in connection with the analysis of meson
decays \cite{Mai:2017vot} and the extraction of scattering phase shifts and resonance parameters
from Lattice QCD \cite{Briceno:2016xwb,Briceno:2017max}. How to formulate an analog of the 
present $N/D$ method for the 3-body system, and how to match the 3-body unitarity formula with
$\chi$EFT calculations, are interesting problems for further study. If our method could be 
extended to the 3-pion cut it would open up applications to the nucleon isoscalar-vector 
and isovector-axial FFs, about which little is known from first principles.
\section*{Acknowledgments}
This material is based upon work supported by the U.S.~Department of Energy, 
Office of Science, Office of Nuclear Physics under contract DE-AC05-06OR23177.
This work was also supported by the Spanish Ministerio de Econom\'ia y Competitividad and
European FEDER funds under Contract No. FPA2016-77313-P.
\appendix 
\section{Expressions}
\label{app:expressions}
For reference we present in this appendix the LO and NLO $\chi$EFT expressions for the
real function $J_+^0(t)$, Eq.~(\ref{J_def}), which are used in the analytical and
numerical studies in the text. 
In the following $4 M_\pi^2 < t < 4 m_N^2$ and [cf.\ Eqs.~(\ref{k_cm}) and (\ref{p_cm})]
\be
k_{\rm cm} = \sqrt{t/4 - M_\pi^2}, \hspace{2em} \widetilde{p}_{\rm cm} = {\textstyle \sqrt{m_N^2 - t/4}}.
\label{k_appendix}
\ee
The contribution of the $N$ Born diagram Fig.~\ref{fig:eft}a is
\ba
J_+^0(t)[\textrm{LO},N] &=& 
\frac{g_A^2 m_N^3}{4\pi f_\pi^2 M_\pi^2} \left( \frac{\arctan x_N}{x_N} \, - \,
\frac{t}{4m_N^2} \right) ,
\hspace{2em}
\label{J_LO_N}
\\[1ex]
x_N &\equiv& \frac{2 k_{\rm cm} \widetilde{p}_{\rm cm}}{A_N} 
\nonumber
\\[1ex]
&=& \frac{2 \, \sqrt{t/4 - M_\pi^2} {\textstyle \sqrt{m_N^2 - t/4}}}{t/2 - M_\pi^2} ,
\label{x_N}
\\[1ex]
A_N &\equiv& t/2 - M_\pi^2 .
\label{A_N_def}
\ea
The contribution of the $\Delta$ Born diagram Fig.~\ref{fig:eft}b is
\ba
J_+^0(t)[\textrm{LO},\Delta]
&=& \frac{h_A^2 }{48 \pi  f_{\pi }^2 M_{\pi }^2} \left[ C_\Delta 
\arctan x_\Delta + D_\Delta \right] ,
\hspace{2em}
\label{J_LO_Delta}
\\[1ex]
x_\Delta &\equiv& \frac{2 k_{\rm cm} \widetilde{p}_{\rm cm}}{A_\Delta} 
\nonumber \\[1ex]
&=&
\frac{2 \, \sqrt{t/4 - M_\pi^2} {\textstyle \sqrt{m_N^2 - t/4}}}{t/2 - M_\pi^2 + m_\Delta^2 - m_N^2} ,
\label{x_Delta}
\\[1ex]
A_\Delta &\equiv& t/2 - M_\pi^2 + m_\Delta^2 - m_N^2 .
\ea
The coefficient of the inverse tangent function in Eq.~(\ref{J_LO_Delta}) is obtained as
\ba
C_\Delta &\equiv& \frac{2 \widetilde{p}_{\rm cm}^2 F - A_\Delta m_N G}{k_{\rm cm} \; \widetilde{p}_{\rm cm}} ,
\label{C_Delta}
\ea
in which
\ba
F &\equiv& \alpha (m_\Delta + m_N) + \frac{\beta}{3} (m_\Delta - m_N) ,
\label{F_def}
\\[1ex] 
G &\equiv& - \alpha + \frac{\beta}{3} ,
\label{G_def}
\\[1ex] 
\alpha &\equiv& \frac{t}{2} - m_N^2 + 
\frac{(m_\Delta^2 + m_N^2 - M_\pi^2)^2}{4 m_\Delta^2} ,
\\[1ex] 
\beta &\equiv& \left( m_N + \frac{m_\Delta^2 + m_N^2 - M_\pi^2}{2 m_\Delta}\right)^2 .
\ea
The full expression for the numerator in Eq.~(\ref{C_Delta}), organized according to powers of 
$M_\pi^2$ and $t$, is
\ba
\lefteqn{2 \widetilde{p}_{\rm cm}^2 F - A_\Delta m_N G \;\; = \;\; 1/(48 m_\Delta^2)} &&
\nonumber \\
&\times&  \left[ 8 m_N (m_N + m_\Delta)^4 (m_N - m_\Delta)^2 \right.
\nonumber \\
&-& 8 m_N (m_N + m_\Delta)^2 (3 m_N^2 - 2 m_N m_\Delta + 3 m_\Delta^2 ) M_\pi^2 
\nonumber \\
&-& 8 m_\Delta  (m_N +m_\Delta)^2 (m_N^2 - 4 m_N m_\Delta + m_\Delta^2 )t 
\nonumber \\
&+& 8 m_N (3 m_N^2 + 2 m_N m_\Delta + 3 m_\Delta^2 ) M_\pi^4 \nonumber
\nonumber \\
&+& 16 m_\Delta (m_N^2 - m_N m_\Delta + m_\Delta^2 ) M_\pi^2 t  
\nonumber \\
&-& \left. 12 m_\Delta^3 t^2  - 8 m_N M_\pi^6 - 8 m_\Delta M_\pi^4 t \right] .
\ea
The polynomial terms in Eq.~(\ref{J_LO_Delta}) are obtained as
\ba
D_\Delta &=& 1/(18 M_\Delta^2) \left[ 6 m_N (m_N + m_\Delta)^3 (m_N - m_\Delta) \right.
\nonumber \\
&+& 4 m_N (4 m_N^2 + 3 m_N m_\Delta + 3 m_\Delta^2) M_\pi^2
\nonumber \\
&-& (19 m_N^3 + 24  m_N^2 m_\Delta + 9 m_N m_\Delta^2 - 6 m_\Delta^3 )t 
\nonumber \\
&-& 6 m_N M_\pi^4 
\; - \; (m_N + 6 m_\Delta ) M_\pi^2 t 
\nonumber \\
&+& \left. (4 m_N + 6 m_\Delta ) t^2 \right] .
\ea

The inverse tangent function in Eqs.~(\ref{J_LO_N}) and (\ref{J_LO_Delta}) contains the logarithmic 
singularity in $t$ resulting from the left-hand cut of the $\pi\pi \rightarrow N \bar N$ PWA.
This singularity corresponds to the intermediate baryon line of the diagrams going on mass shell,
$s = \{ m_N^2, m_\Delta^2 \}$. The coefficient of the singularity is therefore determined
by the $\pi N$ scattering amplitude at the on-shell point. The latter is independent of the
off-shell behavior of the $\chi$EFT even in the case of the intermediate $\Delta$, where the
definition of the $\Delta$ propagator and the $\pi N \Delta$ vertices off the mass shell is
generally ambiguous (for a discussion see Refs.~\cite{Granados:2016jjl,Alarcon:2017asr} 
and references therein). We note that the functions $F$ and $G$ in Eqs.~(\ref{F_def}) and 
(\ref{G_def}) are just the invariant amplitudes of $\pi N$ scattering at $t > 4 M_\pi^2$
and $s = m_\Delta^2$, as defined in Eqs.~(4.15) and (4.16) of
Ref.~\cite{Granados:2016jjl}. The polynomial terms in Eqs.~(\ref{J_LO_N}) and (\ref{J_LO_Delta})
depend on the behavior of the $\pi N$ amplitude off the baryon mass shell. In the case of
the $\Delta$ they depend on the specific choice of the off-shell behavior and the vertices.

The masses and coupling constants used to evaluate the LO expressions are the standard values
for the SU(2) flavor group (see Ref.~\cite{Ledwig:2011cx}): 
$M_\pi = 139\, \textrm{MeV}, f_\pi = 93 \, \textrm{MeV},
m_N = 939\, \textrm{MeV}, g_A = 1.27$, and $m_\Delta = 1232\, \textrm{MeV}, h_A = 2.85$.

The contribution to $J_+^0(t)$ arising from the NLO contact terms in the $\pi N$ amplitude 
is 
\ba
\lefteqn{J_+^0(t)[\textrm{NLO, contact}]
\;\; = \;\; -\frac{\widetilde p_{\rm cm}^2}{12 \pi f_\pi^2 m_N^2 M_\pi^2}
} && \nonumber
\\[1ex]
&\times&
\left[ 12 c_1 M_\pi^2 m_N^2 
+ 2 c_2 \widetilde p_{\rm cm}^2 k_{\rm cm}^2
+ 6 c_3 m_N^2 A_N \right] ,
\hspace{2em}
\label{J_NLO}
\ea
cf.\ Eqs.~(\ref{k_appendix}) and (\ref{A_N_def}). 
The values of the LECs $c_i \; (i = 1, 2, 3)$, determined according to the procedure
described in Sec.~\ref{subsec:higher-order}, are listed in Table~\ref{Tab:LECs}.
%
%
\begin{table}
\begin{ruledtabular}
\begin{tabular}{cccc}
                                & $c_1$ &  $c_2$       &  $c_3$\\
 LECs (GeV$^{-1}$)& (-0.28,-0.18)     &  (1.0, 1.2) & ( -1.64, -0.79)  \\
\end{tabular}
\caption{LECs used in the NLO contact term contribution to $J_+^0(t)$, Eq.~(\ref{J_NLO}).
The values were determined according to the procedure described in Sec.~\ref{subsec:higher-order}.
\label{Tab:LECs}}
\end{ruledtabular}
\end{table}

\end{document}